\documentclass[11pt,cite,epsfig,psfrag]{article}
\usepackage[utf8]{inputenc}
\usepackage{placeins}
\usepackage{geometry}
 \usepackage{graphicx}
 \usepackage{subfig}
  \usepackage{amsmath} 
  \graphicspath{{Images_wgc/}}
  \usepackage[usenames, dvipsnames]{color}
\usepackage{wrapfig}
\usepackage{boxedminipage}
\usepackage{multirow}

\usepackage{fullpage}
\usepackage{amsmath}
\usepackage{amssymb}
\usepackage{graphicx}
\usepackage{mathrsfs}
\usepackage{wrapfig}
\usepackage{boxedminipage}
\usepackage{epsfig}
\usepackage{setspace}
\usepackage{float}
\usepackage{hyperref}
\usepackage{enumerate}
\usepackage{cite}
\usepackage[font=footnotesize,labelfont=rm]{caption}

\usepackage[numbers,sort&compress]{natbib}

\def\be{\begin{equation}}
\def\ee{\end{equation}}
\def\bea{\begin{eqnarray}}
\def\eea{\end{eqnarray}}

\numberwithin{equation}{section}

 \newcommand{\RN}[1]{%
   \textup{\uppercase\expandafter{\romannumeral#1}}%
 }
\begin{document}

\thispagestyle{empty}

\vskip 2cm

\begin{center}
{\Large \bf ISCOs and the weak gravity conjecture bound in higher derivative theories of gravity}
\end{center}

\vskip .2cm

\vskip 1.2cm

\centerline{ \bf   Adrinil Paul \footnote{22ph05002@iitbbs.ac.in} and Chandrasekhar Bhamidipati\footnote{chandrasekhar@iitbbs.ac.in} 
}

\begin{center}{ Department of Physics, School of Basic Sciences\\ 
Indian Institute of Technology Bhubaneswar \\ Bhubaneswar, Odisha, 752050, India}
\end{center}

\vskip 1.2cm
\vskip 1.2cm
\centerline{\bf Abstract}
\vskip 0.5cm

We study circular orbits of charged particles in spherically symmetric AdS black holes in higher derivative theories of gravity, and their limiting ISCOs (innermost stable circular orbits). The dual interpretation is in terms of heavy-light double twist conformal field theory (CFT) operators in the large spin limit, whose anomalous dimensions can be extracted from the binding energy of charged probes in the bulk, in a certain large orbit limit. Demanding the positivity of the anomalous dimensions, leads to an exact bound for the charge to mass ratio $\hat q$ of probe particles in the black hole backgrounds, which matches with the WGC bound. We find that $\hat q$ increases with the higher derivative coupling parameters, which is explicitly checked in the Gauss-Bonnet gravity.  For existing computations with probe particles in AdS backgrounds, the anomalous dimension  and the WGC bound we find, particularly in Gauss-Bonnet theories, are in agreement in appropriate limits with the recent computations for Schwarzschild AdS~\cite{Berenstein:2020vlp}, charged AdS~~\cite{Moitra:2023yyc} and neutral Gauss-Bonnet  black holes in AdS~\cite{Dodelson:2022eiz}. Finally, we show that the ISCOs exist until the limit set by the WGC bound, with their radius decreasing with coupling parameters, which we check explicitly for the case of Gauss-Bonnet black holes in AdS.

\noindent

\vskip 0.5cm
\noindent

\newpage
\setcounter{footnote}{0}
\noindent

\baselineskip 15pt
\section{Introduction}

The weak gravity conjecture (WGC) suggests that any consistent theory of quantum gravity must contain a state whose charge to mass ratio is greater than unity in appropriate units~\cite{Arkani-Hamed:2006emk}. The conjecture has been studied extensively, using both the available theoretical as well as phenomenological framework, in general theories of gravity, due its strong implications for both cosmology and particle physics, including, for the Swampland program in string theory~\cite{Vafa:2005ui} (see also~\cite{Brennan:2017rbf,Palti:2019pca,vanBeest:2021lhn,Harlow:2022ich}). String theory typically contains charged states which satisfy the WGC bound, both above and below the Planck scale as required, such as the extremal black holes in Einstein gravity. With the decrease in black hole mass and the increase of curvature, the higher derivative corrections become important, and extremal black hole charge to mass ratio increases, which is well tested in several scenarios, such as~\cite{Natsuume:1994hd,Kats:2006xp,Cano:2019oma,Cano:2019ycn,Cano:2021nzo,Ma:2021opb} and e.g., from the view point of black hole thermodynamics in~\cite{Cheung:2018cwt,Hamada:2018dde,Bellazzini:2019xts,Charles:2019qqt,Jones:2019nev,Loges:2019jzs,Goon:2019faz,Cremonini:2019wdk,Chen:2020rov,Loges:2020trf,Bobev:2021oku,Arkani-Hamed:2021ajd,Cremonini:2021upd,Aalsma:2021qga,Amsel:2010aj}. In fact, if one extrapolates this extremal charge to mass curve down to the Planck scale, it might lead to a transition from black hole phase to certain microscopic objects, such as excited strings~\cite{Bowick:1985af,Susskind:1993ws,Horowitz:1996nw}, as is known in literature\cite{Heidenreich:2016aqi,Montero:2016tif,Lee:2018urn,Lee:2019tst,Aalsma:2019ryi,Klaewer:2020lfg,Cheung:2014ega,Andriolo:2018lvp,Chen:2019qvr,Alberte:2020bdz,Aoki:2021ckh,Noumi:2021uuv,Noumi:2022zht,Heidenreich:2015nta,Harlow:2022ich}. 

\medskip
\noindent
The study of geodesic motion in black hole backgrounds in the AdS/CFT setting~\cite{Maldacena:1997re,Gubser:1998bc,Witten:1998qj} is quite interesting~\cite{Fidkowski_2004,PhysRevD.79.064016,Kinoshita:2023hgc,Ceplak:2024bja,Hashimoto:2023buz}.  The connection of stable circular orbits in spherically symmetric AdS$_d$ space times (for $d \geq 3$), to the dynamics of certain CFT operators in the dual gauge theory has led to interesting insights in to the anomalous dimensions of CFT operators~\cite{Berenstein:2020vlp,Sekino:2008he,Abajo-Arrastia:2010ajo,Balasubramanian:2011ur,Nozaki:2013wia,Shenker:2013pqa,Festuccia:2008zx,Fitzpatrick:2014vua,Dodelson:2022eiz}.
On the gravity side, stability requires the size of classical orbits to be higher than that of the innermost stable circular orbit (ISCO). Such orbits exist only for $d \geq 3$, and only if their angular momentum $l$ is above the value set by $l_{\rm isco}$ (i.e., the angular momentum of ISCO), otherwise, the orbits fall into the black hole \cite{Cruz:1994ir,Berenstein:2020vlp}. Typically, such plunging orbits correspond to thermalisation, whereas the stable orbits may not thermalise classically. With semiclassical corrections, the ISCOs become metastable to be able to tunnel past the potential barrier, resulting in plunging orbits. Thus, metastable orbits should be interpreted as long lived field theory excitations that do not thermalise on thermal time scales, a fact known from the study of quasinormal modes  in AdS backgrounds. The existence of such orbits can also be understood from dual field theory  point of view, as arising due to the curvature effects of the  theory defined on a sphere, giving an extra scale dependence on the AdS length $L$ (competing with the other scale provided by temperature). Thus, the existence of ISCOs is tied to the global AdS metric. The general conditions which determine the size $r_{\rm isco}$ and the angular momentum $l_{\rm isco}$ of stable orbits are not always solvable analytically, and one has to resort to numerical computations, except in certain extreme limits, such as large mass etc.. In any case, at least for large orbits in the AdS black hole background, it is generally possible to obtain a coordinate independent expression for the binding energies, which can then be transcribed to the CFT language using the usual AdS/CFT dictionary. The resulting expressions are quite illuminative, giving rise to the energies of CFT excitations, which can be obtained independently in the Bootstrap program, from the anomalous dimensions  of composite double twist operators in a large spin limit (though there are some subtleties\cite{Berenstein:2020vlp}). 

\medskip
\noindent
In empty AdS, the boundary interpretation of the orbit states is in terms of a family of certain light double twist operators $[{\mathcal O}_L,{\mathcal O}_L]_{n,J}$, with some spin $J$ and other quantum numbers $n$. Considering one of these operators to be heavy, the interpretation of the heavy-light double twist operators $[{\mathcal O}_H,{\mathcal O}_L]_{n,J}$ in the bulk is in terms of orbit states in the background of a black hole in AdS. Such heavy-light operators have appeared in the study of four point functions in the light-cone bootstrap program. In this regard, the orbit states are expected to be narrow resonances, very similar to the phenomena of quasi-normal modes~\cite{PhysRevD.64.084017}, which can also be understood from the point of view of eigenstate thermalisation hypothesis~\cite{Srednicki_1999,D_Alessio_2016,lashkari2016eigenstatethermalizationhypothesisconformal,10.21468/SciPostPhys.9.3.034}.  In the interest of understanding the dynamics of such metastable states, it is thus important to compute these quantities in general theories of gravity, with additional scalar, and higher derivative corrections. In fact, in~\cite{Berenstein:2020vlp} it was systematically shown how studying the orbits of massive neutral particles in the AdS Schwarzschild black holes in the bulk (and in higher derivative theories~\cite{Dodelson:2022eiz}), provides an understanding of CFT operators, probed earlier using bootstrap techniques~\cite{Fitzpatrick:2014vua}. 

\medskip
\noindent
Studying charged probes in the AdS black hole background, is expected to lead to an improved understanding of anomalous dimensions of heavy-light double twist operators with global $U(1)$ charge on the boundary, particularly, strongly coupled CFTs at large spin as well as charges\cite{Hellerman:2015nra,Alvarez-Gaume:2016vff,Orlando:2019hte,Gaume:2020bmp}. Although, there are considerable differences in the way one takes the large spin and large charge limits of composite CFT operators, under reasonable assumptions, it can be used to study the connection of stable circular orbits with WGC conjecture~\cite{Moitra:2023yyc}.
The set up involving charged particles in charged AdS black hole backgrounds, particularly with the inclusion of cosmological constant, allows an exploration of WGC in the AdS/CFT setting\cite{Nakayama:2015hga,Harlow:2022ich,Aharony:2021mpc,Benjamin:2016fhe,Montero:2016tif,Crisford:2017gsb,Montero:2018fns}. This was in fact exploited recently in ~\cite{Moitra:2023yyc}, where, demanding the positivity of anomalous dimension of certain double-twist operators in the dual CFT (in the large-spin and charge expansion) is found to precisely reproduce the WGC bound for small and large AdS black holes, as well as in the flat-space limit. In addition, it is found that the ISCOs in the AdS black holes stop existing precisely above the limit where the charge to mass ratio saturates the WGC bound. 

\medskip 
\noindent
The subtle interplay of ISCOs and WGC bound should be quite interesting to study in particularly in higher derivative theories of gravity, because there are intriguing corrections to the later, known through independent computations. Since, the ISCOs are expected to be present in general higher derivative theories of gravity, and size $r_{\rm isco}$ as well as the angular momentum $l_{\rm isco}$ of such orbits should be corrected, the connection with WGC needs to be checked from this fresh perspective. Previously, corrections to the WGC bound have been investigated generally from the point of view of self repulsive particles~\cite{Natsuume:1994hd,Kats:2006xp,Cremonini:2009ih,Cano:2019oma,Cano:2019ycn,Cano:2021nzo,Ma:2021opb,Cheung:2018cwt,Hamada:2018dde,Bellazzini:2019xts,Charles:2019qqt,Jones:2019nev,Loges:2019jzs,Goon:2019faz,Cremonini:2019wdk,Chen:2020rov,Loges:2020trf,Bobev:2021oku,Arkani-Hamed:2021ajd,Cremonini:2021upd,Aalsma:2021qga,Amsel:2010aj}. Our interest here is in studying the WGC bound from the computation of binding energies of charged probe particles  propagating in a charged AdS black hole background~\cite{Berenstein:2020vlp,Moitra:2023yyc,Dodelson:2022eiz}, in the presence of certain higher derivative terms in the action~\cite{PhysRevLett.55.2656, Nojiri_2001, NOJIRI2006144, Cai:2001dz, Cveti__2002,Cai:2013qga}. Specifically, we consider charged black holes in AdS in Gauss-Bonnet theory of gravity, which contain extremal black holes, that describe zero temperature CFTs at finite chemical potential. This set up of course has been studied extensively in the holographic context, with applications to condensed matter systems and exploring phenomena such as, transport coefficients and holographic superconductors~\cite{Cai:2009zn,Cai:2010zm}. Exact analytical expressions for energy and angular momentum of charged particles in large stable circular orbits, in arbitrary dimensions can be written down. The dual interpretation should again be in terms of certain heavy-light double twist operators with large spin and also charge, whose anomalous dimensions could be extracted from the energy of bulk particle probes (as done earlier for neutral probes in~\cite{Dodelson:2022eiz}).  In addition, it is known that there are small effects (at large ’t Hooft coupling) on the Regge trajectories in the presence of higher derivative corrections~\cite{Camanho_2016,Afkhami_Jeddi_2017, afkhamijeddi2017shockwavesoperatorproductexpansion,Kulaxizi_2018, Costa_2017,Caron_Huot_2023,Grinberg_2021}, apart from the simplifications that occur in the conformal bootstrap analysis in the large spin limit~\cite{Alday:2007mf,Komargodski:2012ek,Alday:2015eya,Kulaxizi:2018dxo, Karlsson:2019qfi,Li:2019zba,Li:2020dqm}. For charged probes, such an analysis, then leads us to the WGC bound  depending on the Gauss-Bonnet coupling parameter $\alpha$, which are consistent with the earlier computations.  For existing computations with probe particles in AdS backgrounds, the anomalous dimension  and the WGC bound we find, particularly in Gauss-Bonnet theories, are in agreement in appropriate limits with the recent computations for Schwarzschild AdS~\cite{Berenstein:2020vlp}, charged AdS~~\cite{Moitra:2023yyc} and neutral Gauss-Bonnet  black holes in AdS~\cite{Dodelson:2022eiz}. We find that the WGC bound coming from the charge to mass ratio of particles in circular orbits in arbitrary dimensions, increases with the $\alpha$, in agreement with earlier computations~\cite{Kats:2006xp,Cremonini:2009ih,Dodelson:2022eiz}, though there are  disagreements with~\cite{Kats:2006xp,Cremonini:2009ih}. In general, the series expansion of WGC bound in terms of higher derivative couplings, found from probe particles in AdS black hole backgrounds, does not match the corresponding expressions known for self-repulsive particles.  We also compute the radius of ISCOs of charged particles in the AdS black holes in Gauss-Bonnet AdS black holes, and show that the ISCOs stop existing precisely when the WGC bound is satisfied. Although, the radius of ISCOs cannot be computed analytically, the numerical analysis shows a suppression of this radius with $\alpha$. For neutral probes, the reduction in $r_{\rm isco}$ can be seen analytically, going below that of the AdS Schwarzschild value~\cite{Berenstein:2020vlp}.

\medskip
\noindent
Gauss-Bonnet terms alluded to above are only one set of higher derivative terms and in general, the low energy effective action is expected to have other four derivative terms, with coupling between gauge and gravity fields. Following the set up in~\cite{Cremonini:2019wdk}, we study the energy and angular momentum of charged probes in black holes in AdS in higher derivative theories, which depend on several Wilson coefficients. In certain specific limits, we obtain the coordinate independent expressions for binding energy of charged probes, which via AdS/CFT gives rise to expression for anomalous dimension of corresponding CFT operators. The positivity of the anomalous dimension leads to a bound, which leads us to the WGC bound in higher derivative theories.

\medskip
\noindent
Summary and plan for rest of the paper is as follows. In section-(\ref{two}), we derive the d-dimensional expressions for energy and angular momentum of charged particles in a charged Gauss-Bonnet AdS black hole. The binding energy of the particles, in particular, is expressed in a coordinate independent manner in certain large radius limit, in terms of the effective AdS length scale $L_{\rm eff}$ in eqn. (\ref{be}). We then use AdS/CFT correspondence, to express the binding energy of charged particle motion in small black holes in terms of CFT quantities, which gives the anomalous dimension of boundary operators in eqn. (\ref{ana}), which is found to be sign indefinite, as in~\cite{Moitra:2023yyc}. The WGC bound follows from demanding positivity of anomalous dimensions and is given in eqn. (\ref{wgc}). We then extend the results to case of more general case of black holes in four  derivative theories~\cite{Cremonini:2019wdk}, and obtain a WGC bound given in eqn. (\ref{wgc2}) of section-(\ref{three2}). Section-(\ref{three}) contains a computation of radius of ISCO, which is shown to decrease with the Gauss-Bonnet coupling parameter $\alpha$. Numerical plots show that the ISCOs stop existing precisely when the increased charge to mass ratio and the WGC bound is met. We end with conclusions in section-(\ref{four}).

\section{Binding energy of charged probes and WGC } \label{two}
In the following subsection-(\ref{2.1}), we derive the general d-dimensional expressions for energy and angular momentum of charged particles in a charged Gauss-Bonnet AdS black hole, and spell out their forms in the small and the large black hole limits. This is then used to obtain the coordinate independent expression for binding energy of charged probes for large orbits, in subsection-(\ref{2.2}). The anomalous dimensions of the possible dual double twist operators in the large spin (and charge) limit, can be extracted from the binding energy, which also leads us to the WGC bound. We repeat the above exercise in subsection-(\ref{three2}) when the action contains more general four derivative terms. 

\subsection{Circular orbits in Gauss-Bonnet AdS black holes} \label{2.1}
We start with the action of the Einstein- Maxwell action with a cosmological constant and a Gauss-Bonnet term in $d-$ dimensional spacetimes taken as~\cite{Cai:2001dz,Cai:2013qga}, 
\begin{equation}
    S=\frac1{16\pi}\int d^dx \sqrt{-g}[R-2\Lambda+\alpha (R_{\mu\nu\gamma\delta}R^{\mu\nu\gamma\delta}-4R_{\mu\nu}R^{\mu\nu}+R^2)-4\pi F_{\mu\nu}F^{\mu\nu}] \, ,
\end{equation}
where $\alpha$ is the Gauss-Bonnet coefficient with dimension $[{\rm length}]^2$.  The cosmological constant is $\Lambda=-\frac{(d-1)(d-2)}{2L^2}$, where  $F_{\mu\nu}=\partial_\mu A_\nu-\partial_\nu A_\mu$ is the Maxwell field strength with vector potential $A_\mu$.
The action is known to admit a static spherically symmetric black hole solution with metric,
\begin{equation} \label{GBm}
ds^2=-f(r)dt^2+f^{-1}(r)dr^2+r^2d\Omega_k,
\end{equation}
 where $d\Omega_k$ here represents the line element of a $(d-2)$-dimensional maximally symmetric Einstein space having constant curvature $(d-2)(d-3)$k, with volume $\Sigma_k$ as
 \begin{equation}
     \Sigma_k= \frac{2\pi^{[\frac{D-1}{2}]}}{\Gamma[\frac{D-1}{2}]} \, .
 \end{equation}
The function $f(r)$ is,
\begin{equation}
f(r)=1+\frac{r^2}{2\widetilde{\alpha}}\left (1-\sqrt{1+\frac{4\widetilde{\alpha} M}{r^{d-1}}-\frac{4 \widetilde{\alpha} Q^2}{r^{2d-4}}+\frac{8\widetilde{\alpha} \Lambda}{(d-1)(d-2)}} \right ),\label{4}
\end{equation}
where $\widetilde{\alpha}=(d-3)(d-4)\alpha$. The black hole mass is $M=\frac{16\pi M_{BH}}{(d-2)\Sigma_k}$ with $Q^2=\frac{ Q_{BH}^2}{2(d-2)(d-3)}$ denoting the charge of the black hole. The electric potential of the black hole in $d$-dimension is given as,
\begin{equation}
   A_t(r)=-k\frac{Q}{r^{d-3}} \, , \label{5}
\end{equation} 
where the constant $k=\sqrt{\frac{(d-2)}{2(d-3)}}$. 
The horizon radius $r_h$  of the black hole is determined by the largest real root of the equation $f(r_h)=0$ and the mass  $M$ can be expressed in terms of the horizon $r_h$ as,
\begin{equation}
M=\frac{Q^2}{r_h^{d-3}}+r_h^{d-3}\left(1+\frac{r_h^2}{L^2}\right)+\frac{\tilde{\alpha}}{r_h^{d-5}} \, .
\end{equation}
We can obtain the extremal mass and charge in terms of horizon radius by setting $f(r_h)=f'(r_h)=0$ to be, respectively,
\begin{equation}
    M_{ext}= \frac{2 r_h^{-5+d} \left((-3+d) L^2 r_h^2+(-2+d) r_h^4+(-4+d) L^2 \tilde{\alpha }\right)}{(-3+d) L^2} \, ,\label{7}
\end{equation}
\begin{equation} \label{Qext}
    Q_{ext}= \frac{ r_h^{-4+d} \sqrt{(-3+d) L^2 r_h^2+(-1+d) r_h^4+(-5+d) L^2 \tilde{\alpha }}}{\sqrt{-3+d} L} \, .
\end{equation}
The idea is to obtain the orbits of a probe particle having charge to mass ratio $\hat q = q/m$ in the above background geometry. The motion of such particles can concisely be described by a one-dimensional effective potential
\begin{equation}
    V(r)=\left(1+\frac{\hat l^2}{r^2}\right)f(r)-(\hat E+\hat qA_t)^2  \, \label{9}
\end{equation}
As is known, there are two conserved quantities associated with the motion of such particles, namely, the energy and angular momentum, denoted respectively as $\hat E = E/m$ and $\hat l = l/m$. Before writing the expressions, it is useful to consider the case of global AdS $(M=Q=0)$. This is important because for fixed $M$ and $Q$, large circular orbits should approach the values in global AdS case. The critical points can be easily found and we get: 
\begin{equation}
r^2= \hat l \, L_{\rm eff}\, ,
\end{equation}
where we defined for convenience
\begin{equation} \label{effe}
     L_{\rm eff}^2=\frac{2\widetilde{\alpha}}{1-\sqrt{1-\frac{4\widetilde{\alpha}}{L^2}}} \, ,
 \end{equation}
 remembering that asymptotically, the geometry is AdS with the usual length replaced by the expression in eqn. (\ref{effe})~\cite{Berenstein:2020vlp}. 
This relation is also useful when comparing with CFT quantities in the computation of anomalous dimensions in eqn. \eqref{cft op}.

\medskip
\noindent
Now, the expressions for energy and angular momentum in the background eqn. (\ref{GBm}) can be obtained straightforwardly by evaluating the conditions $V(r_c) = 0  = V'(r_c)$ for circular orbits of radius $r_c$ and given as 
{\small	\begin{equation} \label{E1}
   \hat E= \frac{k \hat q Q}{ r^{d-3}}-\frac{r^d A \left(r^d \sqrt{4+\frac{(d-3)^2 k^2 \hat q^2 Q^2}{ r^{2d-6}}+\frac{ 2 B}{ r^{2d-3}A}}-(d-3) k \hat q Q r^3\right) \left(-2 \tilde{\alpha }+r^2 \left(-1+A\right)\right)}{2 \tilde{\alpha } \left(2(d-2) Q^2 r^6-(d-1) M r^{3+d}+2 r^{2 d} A\right)}\, ,
\end{equation}}
{\small \begin{equation} \label{l1}
   \hat l^2=-r^2 \left(-1+\frac{\left(-L^2 r^{2 d}+4 C \tilde{\alpha }\right) \left(r^d \sqrt{4+\frac{(d-3)^2 k^2 \hat q^2 Q^2}{ r^{2d-6}}+\frac{ 2 B}{ r^{2d-3}A}}-(d-3)
k \hat q Q r^3\right)^2 \left(-2 \tilde{\alpha }+r^2 \left(-1+A\right)\right)}{2 L^2 \tilde{\alpha } \left(2(d-2) Q^2 r^6-(d-1) M r^{d+3}+2 r^{2 d} A\right)^2}\right)\, ,
\end{equation}}
with,
{\small 
\begin{equation*}
   A= \sqrt{1+\left(-\frac{4}{L^2}-4 Q^2 r^{4-2 d}+4 M r^{1-d}\right) \tilde{\alpha }} \, , 
\end{equation*}}
\begin{equation*}
     B=\left(2(-2+d) Q^2 r^3-(-1+d) M r^d\right) \,,
\end{equation*}
\begin{equation}
    C=\left(r^{2 d}+L^2 r \left(Q^2 r^3- M r^d\right)\right)\, .
\end{equation}
It is understood that one evaluates the functions and derivatives in the above expressions at $r=r_c$. The situation in the asymptotically flat limit is analogous to that found in~\cite{Moitra:2023yyc} and we do not repeat it here. It is important to remember that in this limit, large circular orbits exist only for $M > k_d \hat q Q$ and are unstable for $d \geq 5$, with stability occurring only in the special dimension $d=4$, particularly, because of the interplay of Newtonian and Coulomb potentials. In the presence of cosmological constant, we first study the dynamics of charged test particles around a small black hole $(r_h \ll L)$  moving in a fairly large circular orbit, keeping in mind that the orbit radius is much smaller than the AdS-radius. The condition which describes this is, 
\begin{equation}
    \frac{M}{r^{d-3}} \ll \frac{r^2}{L^2} \ll 1 \, . \label{cond}
\end{equation}
In this regime, energy and angular momentum take the form,
\begin{equation*}
 \hat    E=1+\frac{r^2}{L_{\rm eff}^2}+ \frac{(d-5)}{4 r^{d-3}}\left(\frac{ML_{\rm eff}^2}{L_{\rm eff}^2-2\widetilde{\alpha}}-2k \hat qQ\right) \, ,
\end{equation*}
\begin{equation}
 \hat       l^2=\frac{r^4}{L_{\rm eff}^2}\left(1+L_{\rm eff}^2\frac{(d-3)}{2r^{d-1} }\left(\frac{ML_{\rm eff}^2}{L_{\rm eff}^2-2\widetilde{\alpha}}-2k \hat qQ\right)\right) \, .\label{ELs}
\end{equation} 
We also record the condition for large black holes for later use, obtained as,
\begin{equation}
    \frac{M}{r^{d-3}} \ll 1 \quad \& \quad \frac{r^2}{L^2} \gg 1  \label{cond}
\end{equation}
Of course, since the circular orbit radius is always greater than horizon radius, the above condition actually is valid for black holes of all the sizes in AdS. 
In this regime, energy and angular momentum  take the form,
\begin{equation*}
 \hat      E=1+\frac{r^2}{L_{\rm eff}^2}\left(1+\frac{(d-1)}{4}\left(\frac{L_{\rm eff}^2}{L_{\rm eff}^2-2\widetilde{\alpha}}\right)\frac{M}{r^{d-3}}-\frac{(d-3)k \hat qQ}{2r^{d-3}}\right)+\frac{(d-5)}{4 r^{d-3}}\left(\frac{ML_{\rm eff}^2}{L_{\rm eff}^2-2\widetilde{\alpha}}-2k \hat qQ\right) \, ,
\end{equation*}
\begin{equation}
   \hat    l^2=\frac{r^4}{L_{\rm eff}^2}\left(1+\frac{(d-1)}{2}\left(\frac{L_{\rm eff}^2}{L_{\rm eff}^2-2\widetilde{\alpha}}\right)\frac{M}{r^{d-3}}-\frac{(d-3)k \hat qQ}{r^{d-3}}+L_{\rm eff}^2\frac{(d-3)}{2r^{d-1} }\left(\frac{ML_{\rm eff}^2}{L_{\rm eff}^2-2\widetilde{\alpha}}-2k \hat qQ\right)\right) \, .\label{ Large BH E and l}
    \end{equation}
    
\subsubsection{WGC bound from anomalous dimensions of double-twist operators} \label{2.2}
\noindent
We now look to extract the anomalous dimension of charged CFT operators by obtaining a coordinate invariant expression for the binding energy of probe particles. This will also lead us to an exact expression for the WGC bound in the Gauss-Bonnet theory, which we try to match with similar computations in various limits from~\cite{Berenstein:2020vlp,Dodelson:2022eiz,Moitra:2023yyc}. WGC bound in higher derivative theories of gravity has been computed earlier in literature from a different point of view, and one such set up involves the case where the higher derivative terms are treated perturbatively, while keeping the black hole solution at the two derivative level exact. This is the situation involving self-repulsive particles~\cite{Kats:2006xp,Cremonini:2009ih}. Though, it is not clear whether a direct comparison of our results is possible with this set up, we still go ahead and try to match the bound we obtained within the AdS/CFT setting with perturbative series expansions in terms of higher derivative couplings obtained in~\cite{Kats:2006xp,Cremonini:2009ih} and find disagreements.

\medskip
\noindent
Considering the small black hole regime for the moment, one can now express the energy entirely in terms of angular momentum and other parameters, by eliminating the radius $r_c$ (i.e., by solving eqn. (\ref{ELs} perturbatively~\cite{Berenstein:2020vlp,Dodelson:2022eiz,Moitra:2023yyc}) as,
\begin{equation}
  \hat   E=1+\frac{ \hat   l}{L_{\rm eff}}-\left(\frac{ML_{\rm eff}^2}{2(L_{\rm eff}^2-2\widetilde{\alpha})}-k \hat qQ\right)\left(\frac{1}{ \hat  l L_{\rm eff}}\right)^\frac{d-3}{2} \label{be}
\end{equation}
\noindent
The above equation is valid for the large black hole as well. 
With the above set up, the expression in eqn. (\ref{be}) for binding energy can now be translated into the CFT quantities. As per the AdS/CFT dictionary, considering a scalar field with mass $m$ (and charge $q$) in the bulk (AdS$_{d+1}$), corresponds to an operator on the boundary with the dimension $\Delta_q = d/2 + \sqrt{m^2 L^2 + (d/2)^2}$. In the limit $m \gg 1/L$, one can take $\Delta_q \approx m L$ (assuming $\Delta_q \gg 1$), the description of the bulk probe effectively becomes classical, and the single particle states in the bulk get mapped to single trace operators on the boundary. For the present situation, following\cite{Berenstein:2020vlp, Dodelson:2022eiz} and using $\hat l \to \mathcal{J} (L_{\rm eff}/\Delta_q)$ as well as $\hat q \equiv \hat q (L_{\rm eff}/ \Delta_q)$ (with $\mathcal{J}$ and $q$ being respectively the angular momentum and charge of the boundary operator), the Regge trajectory for large $\mathcal{J}$ is~\cite{Berenstein:2020vlp,Dodelson:2022eiz},
\bea
&& \Delta_{Q,q} (\mathcal{J}) = \Delta_Q + \Delta_q + \mathcal{J} + \gamma (M, Q, \hat q , \mathcal{J}, \alpha)\, .  \label{regge}
\eea
Here, the $\Delta_Q, \Delta_q $ correspond respectively to rest energies of the black hole and the probe particle.   $\mathcal{J}$ is the contribution of spin to the energy,  and  $\gamma$ is the anomalous dimension, which represents the binding energy. Compared to~\cite{Berenstein:2020vlp,Dodelson:2022eiz}, the changes are the modification of the bare AdS radius $L$ to $L_{\rm eff}$, together with a dependence of the anomalous dimension $\gamma$ on $\alpha$, which are expected in this theory.
\medskip
\noindent
Now, using the above dictionary, to write the energy $\hat E$ and angular momentum $\hat l$ per unit mass of the bulk particle, and translate into an energy $\mathcal{E}$ and angular momentum $\mathcal{J}$ in the CFT, we get
\begin{equation}
\mathcal{E}=\Delta_q \hat E,  \quad \quad   \mathcal{J}=\Delta_q\frac{ \hat l}{L_{\rm eff}} \label{cft op}
\end{equation} 
Using eqn.\eqref{cft op} we can map eqn.\eqref{be} to the required CFT expression,
\begin{equation}
    \mathcal{E}=\Delta_q+\mathcal{J}-\Delta_q\left(\frac{ML_{\rm eff}^2}{2(L_{\rm eff}^2-2\widetilde{\alpha})}-k \hat qQ\right)\left(\frac{1}{L_{\rm eff}}\right)^{d-3}\left(\frac{\Delta_q}{\mathcal{J}}\right)^{\frac{d-3}{2}}
\end{equation}
Comparing with equation (\ref{regge}), one writes
{\large\begin{equation}
    \gamma=\Delta_q( \hat E- \hat l-1) |_{\hat l=\frac{\mathcal{J}L_{\rm eff}}{\Delta_q}} \, ,
\end{equation}}
and the anomalous dimension $\gamma$ is therefore
\begin{equation} \label{ana}
   \gamma=-\Delta_q\left(\frac{ML_{\rm eff}^2}{2(L_{\rm eff}^2-2\widetilde{\alpha})}-k \hat qQ\right)\left(\frac{1}{L_{\rm eff}}\right)^{d-3}\left(\frac{\Delta_q}{\mathcal{J}}\right)^{\frac{d-3}{2}} \, .
\end{equation}
Few comments regarding matching eqn. (\ref{ana}) with earlier results in certain limits are in order:
\begin{itemize}
\item
In the limit $ \alpha \rightarrow 0, Q \neq 0$,  we have $L_{\rm eff}=L$ and the above expression in eqn. (\ref{ana}) exactly matches the one for charged black holes in AdS computed earlier\footnote{While matching with~\cite{Moitra:2023yyc}, we note that $f(r)$ in eqn. (\ref{4}) goes over to the corresponding expression for charged black holes in AdS, in the limit $\alpha \rightarrow 0$, after scaling our mass $M$ by a factor of 2.} (see eqn.(17) of~\cite{Moitra:2023yyc}).

\item
In the limit $ \alpha \rightarrow 0, Q =0$, the results for the case of Schwarzschild black holes in AdS can be recovered as well (see the last term for anomalous dimensions in eqn. (57) of ~\cite{Berenstein:2020vlp} ).

\item
In the limit $Q =0, \alpha \neq 0$, which is the case of neutral Gauss-Bonnet black holes in AdS, eqn. (\ref{ana}) can be expanded in a series as
\begin{equation}
    \gamma = -\Delta\left(\frac{M}{L^{d-3}}+\alpha\frac{d+1}{2L^{d-1}}M+\alpha^2\frac{(d+5)(d+3)}{8L^{d+1}}M +\mathcal{O}(\alpha^3)\right)\left(\frac{\Delta}{J}\right)^\frac{d-3}{2}\, .
\end{equation}
We thus find an increase in the anomalous dimension with the Gauss-Bonnet coupling $\alpha$ (assuming this parameter is positive as is known from heterotic string theory, and meeting the bounds in~\cite{PhysRevD.77.126006}). This result was seen earlier in a perturbative computation of anomalous dimensions in the small mass limit (see eqn. (2.30) of~\cite{Dodelson:2022eiz}).

\end{itemize}

\medskip
\noindent
The anomalous dimension in eqn. (\ref{ana}) is sign-indefinite and alerts to a bound. For a black hole of charge $Q=Q_{\rm ext}$, the minimum mass will be $M=M_{\rm ext}$. Therefore the anomalous dimension of like charged CFT operator is positive if,
\begin{equation}
\hat     q \ge \hat q_0  \left(\frac{L_{\rm eff}^2}{L_{\rm eff}^2-2\tilde{\alpha}}\right)\frac{M_{\rm ext}}{Q_{\rm ext}} \, , \label{wgc}
\end{equation}
where $\hat q_0 = \sqrt{\frac{(d-3)}{2(d-2)}} $.
We have the following observations:
\begin{itemize}
\item
The above bound generalizes the one obtained earlier for Charged black holes in AdS and agrees with it in the limit $\alpha \rightarrow 0 $ (see eqn.(18) of ~\cite{Moitra:2023yyc}). Eqn. (\ref{wgc}) is also the WGC bound in the Gauss-Bonnet theory, obtained independently with in the AdS/CFT setting. 
\item
In the limit $r<<L$, which is the flat space-time limt, and taking $d=5$, one gets the following series from eqn. (\ref{wgc}),
\begin{equation} \label{flat1}
\hat q \ge \hat q_0 \Big|_{d=5} \left(1+ \frac{\alpha}{r^2}  \right) + \cdots \, .
\end{equation}
We try to match our above result with existing results~\cite{Kats:2006xp,Cremonini:2009ih}, and two comments are in order. First, as long as $\alpha$ is positive, one expects the higher derivative corrections to enhance the charge to mass ratio (independent of the number of space-time dimensions), which is one of the predictions of WGC, and observed earlier in a different setting~\cite{Kats:2006xp}(see also eqn. (66) of ~\cite{Cremonini:2009ih}). Second, in general dimensions, one expects the factor $\hat q_0$ in eqn. (\ref{wgc}) (or in eqn. (\ref{flat1})) to be the leading order result. The inverse of this factor, i.e., $1/{\hat q_0}$, gives the leading order mass to charge ratio, which matches with eqn. (67) of~\cite{Cremonini:2009ih}. This might be a coincidence in flat space-time, as the rest of the terms in eqn. (\ref{flat1}) do not match with the corresponding expressions in eqn. (66) of~\cite{Cremonini:2009ih}.
\item
In the small $\alpha $ limit in five dimensions, we get
\begin{equation} \label{bound}
\hat q \ge \left( \frac{3 \beta +2}{\sqrt{6 \beta +3}}\right) \left( 1+ 2 \alpha  \frac{ \left(6 \beta ^2+4 \beta
   +1\right)}{\beta  (3 \beta +2) L^2}\right) \, + O(\alpha^2), \quad \text{where} \quad  \beta = \frac{r_h^2}{L^2}
\end{equation}
showing enhancement in charge to mass ratio with $\alpha$. A series for mass to charge ratio of self-respulsive particles was computed in ~\cite{Cremonini:2009ih}, where the higher four derivative terms are considered as perturbations to the two derivative results. 
We find that our above series in eqn. (\ref{bound}) for probe particles in a black hole background, does not match the corresponding expression given in eqn. (A5) of ~\cite{Cremonini:2009ih}.

\end{itemize}
\noindent
We thus find that the exact WGC bound in the Gauss-Bonnet gravity in eqn. (\ref{wgc}) for particle probes in a black hole background, when expanded as a series (either in the small $\alpha$ limit or the flat space-time limit), does not match the earlier results found from the computations done for self-repulsive particles in~\cite{Natsuume:1994hd,Kats:2006xp,Cano:2019oma,Cano:2019ycn,Cano:2021nzo,Ma:2021opb,Cheung:2018cwt,Hamada:2018dde,Bellazzini:2019xts,Charles:2019qqt,Jones:2019nev,Loges:2019jzs,Goon:2019faz,Cremonini:2019wdk,Chen:2020rov,Loges:2020trf,Bobev:2021oku,Arkani-Hamed:2021ajd,Cremonini:2021upd,Aalsma:2021qga,Amsel:2010aj}. We should however emphasise that:  for situations where the computations exist for the case of probe particles in AdS backgrounds, our results for the anomalous dimension in eqn. (\ref{ana}) and/or the WGC bound in eqn. (\ref{wgc}) are in agreement in appropriate limits with the recent computations for Schwarzschild AdS~\cite{Berenstein:2020vlp}, charged AdS~\cite{Moitra:2023yyc} and neutral Gauss-Bonnet  black holes in AdS~\cite{Dodelson:2022eiz}.

\medskip
\noindent
Thus far, we considered small black holes $(r_h \ll L)$. We can extrapolate the WGC Bound for large black hole$(r_h \gg L)$ and possibly for black holes of all sizes, following the arguments in~\cite{Moitra:2023yyc}. We skip the details and instead note that in this limit we get, 
\begin{equation}
   \hat q \ge \sqrt{\frac{2}{d-2}} \left(\frac{(d-2)\frac{r_h^2}{L^2}+(d-4)\frac{\tilde \alpha}{r_h^2}}{\sqrt{(d-1)\frac{r_h^2}{L^2}+(d-5)\frac{\tilde \alpha}{r_h^2}}}\right)\left(\frac{L_{\rm eff}^2}{L_{\rm eff}^2-2\tilde{\alpha}}\right)
\end{equation}
In $d=5$, for small $\widetilde{\alpha}$,
\begin{equation}
\hat    q \ge \sqrt{\frac{3}{2}} \frac{r_h}{L}+ 2 \sqrt{6}\frac{r_h}{L^3} \alpha+O[\alpha^2 ]
\end{equation}  
With appropriate identifications, the leading term of the above expression agrees with eqn. (19) of~\cite{Moitra:2023yyc}.


\subsection{WGC bound in four-derivative theories} \label{three2}
Encouraged by the appearance of a WGC bound for the Gauss-Bonnet black holes in eqn. (\ref{wgc}), we seek to extend the results to a situation when the Einstein-Maxwell theory with a negative cosmological constant, is supplemented by more general four-derivative terms. For this purpose, the 
set up of~\cite{Cremonini:2019wdk} is useful, where higher derivative corrections to extremal black hole mass and charge were evaluated perturbatively, with the action taken as:
\begin{align}
    \begin{split}
        I &= -\frac{1}{16 \pi } \int d^{d} x \sqrt{-g} \Bigg[  \frac{(d-2) (d-1)}{L^2} + R -\frac{1}{4} F^2 \\
        & \qquad \qquad \qquad \qquad +L^2  \epsilon \Big( c_1 R_{abcd} R^{abcd} +  c_2 R_{abcd} F^{ab} F^{cd} + c_3 (F^2)^2 + c_4 F^4 \Big) \Bigg].
    \end{split}
    \label{Action}
\end{align}
Turning off the higher derivative corrections (by setting the book keeping parameter $\epsilon$ to zero), the action will admit charged-AdS black hole solution following from the two derivative theory. As the equations of motion are not easy to solve in general, the usual procedure is to freeze the black hole solution found in the two derivative theory, and obtain the corrections to the metric perturbatively, by finding the effect of higher derivative terms through boundary stress tensor. 
The first order solution obtained this way is~\cite{Cremonini:2019wdk}
\begin{align} \label{metricRF}
    &ds^2=-f(r)dt^2+g(r)^{-1}dr^2+r^2d\Omega_{d-2}^2 \, ,
\end{align}
where,
\begin{align}
    f(r) = (1 + \gamma(r)) g(r),
\end{align}
with,
\begin{align}
    \gamma(r) =\left( c_1 \frac{(d - 3)(2 d^2 - 9 d + 8) }{(d - 2) } + c_2 (d-1) (d - 3)\right)\frac{Q^2L^2}{r^{2d-4}}.
    \label{eq: gammaVal}
\end{align}
and
\begin{equation}
    g(r)=1-\frac{M}{r^{d-3}}+\frac{Q^2}{4r^{2d-6}}+\frac{r^2}{L^2}+\Delta g.
\end{equation}
with
\begin{align}
    \begin{split}
        &\Delta g (r) = \\ 
        &c_1 \Bigg( -\frac{ (d - 3) (8 d^3 - 48 d^2 + 87 d -44) \, Q^4 L^2}{ 8 (d - 2) (3 d - 7) r^{4d - 10}}
        + \frac{(d - 3) (4 d^2 - 17 d + 16 ) \, M Q^2 L^2 }{2 (d - 2) r^{3d - 7}} 
        \\
        & \qquad  -  \frac{4  (d-3)^2 \, L^2 Q^2}{r^{2d-4}}
        + \frac{  (d-3) (d-4) \, L^2 M^2}{r^{2d-4}}
        - \frac{ (2d - 5)(2 d^2 - 9 d + 8) \, Q^2 }{(d - 2)r^{2 d-6}}
        + \frac{2 (d - 4) r^2}{(d - 2)L^2} \Bigg)
        \\
        & + c_2 \Bigg( - \frac{ (d-2)^2 (d-3) \, Q^4 L^2}{(3d - 7)r^{4d - 10}}
        + \frac{ (3d^2 - 14 d + 15) \, Q^2 M L^2}{2 r^{3d-7}}
        -  \frac{2 (d-2) (d-3) \, Q^2 L^2}{r^{2d-4}} \\
        &\hspace{12mm}- \frac{2 (d-2)^2 \, Q^2}{r^{2d-6}} \Bigg)  + \left( 2 c_3 + c_4 \right) \Bigg(- \frac{(d - 2) (d - 3)^2 Q^4 L^2 }{ (6d - 14) r^{4d - 10}} \Bigg) \, .
    \end{split}
    \label{eq:Deltag}
\end{align}
The functions $\Delta g$ and $\gamma(r)$ depend on the Wilson coefficients $c_i$ and encode the effect of higher derivative terms. Finally, the full electric field is,
\begin{align}
    \begin{split}
        F_{tr} &= \sqrt{\frac{(d - 3)(d - 2) }{2}} \Bigg[(1-8c_2) \frac{Q}{r^{d - 2}}+4c_2(d-2)(d-3)\frac{QML^2}{r^{2d-3}}\\
        &\kern4em+\left( \frac{c_1}2\frac{(2 d^2 - 9 d + 8)}{(d - 2)} -\frac{ c_2}2 (7d - 19) - 4\left( 2 c_3 + c_4 \right)(d - 2) \right) (d-3)\frac{Q^3L^2}{r^{3d-6}}
        \Bigg] \, ,
    \end{split}
    \label{eq:Ftr}
\end{align}
Let us briefly summarize the method used in ~\cite{Cremonini:2019wdk} for obtaining the mass to charge ratio of extremal black holes. First one computes the extremal mass $M_0$ and charge $Q_0$ of the two derivative theory exactly. Then, one directly writes the higher derivative corrected expressions $M$ and charge $Q$ by manually adding the contribution from various functions of the radial coordinate, given in $\Delta g$ (see eqn. (3.2) of~\cite{Cremonini:2019wdk}). Thus, the mass to charge ratio $M/Q$ is a series in the Wilson coefficients with the leading term being $M_0/Q_0$, though one has to hold some quantities of the two derivative theory fixed (such as, mass or charge or the horizon radius), while the other quantities get the corrections (see e.g., eqn. (3.12) of~\cite{Cremonini:2019wdk} and the discussion there in). Though the details of the result for $M/Q$ differ in each case, the common feature is that this ratio decreases with the Wilson coefficients, irrespective of which of the quantities of the two-derivative theory are held fixed. \\

\noindent
Our strategy and computation below differs from the above methods. First, we work with the full metric in eqn. (\ref{metricRF}) and actually set $\epsilon=1$, so that all the terms in the functions $\gamma(r)$ and $g(r)$ are taken exactly in the computation of energy and angular momentum of test particles (see appendix-(\ref{AppendixA})) in the black hole background in eqn. (\ref{metricRF}). 
We compute the charge to mass ratio, i.e., $\hat q = q/m$ instead, for the probe particles in the black hole background (and not that of the black hole or self-repulsive particle itself). The result for the WGC bound we find is exact in a higher derivative theory, with no quantities held fixed. Thus, one needs to be prudent in making a direct comparison with the bound studied in~\cite{Cremonini:2019wdk}.\\

\noindent
The method for obtaining the orbits of a probe particles having charge to mass ratio $\hat q = q/m$ in the background geometry of eqn. (\ref{metricRF}) is same as before with the effective potential given in eqn. (\ref{9}). As before, there are two conserved quantities associated with the motion of such particles, namely, the energy and angular momentum, denoted respectively as $\hat E = E/m$ and $\hat l = l/m$. Large circular orbits in the black hole background should approach the values known in global AdS case, for which the circular orbit radius can be found to be $r^2= \hat l \, L_{\rm eff}$,
where we defined for convenience
 \begin{align} \label{effe1}
    L^2_{\text{eff}} = L^2/\lambda^2 , \qquad \qquad \lambda^2 = \left( 1 + c_1 \frac{2 (d - 4)}{(d - 2)} \right).
\end{align}
 remembering that asymptotically, the geometry is AdS with the usual length replaced by the expression in eqn. (\ref{effe1})~\cite{Berenstein:2020vlp,Cremonini:2019wdk}. 
This relation is also useful when comparing with CFT quantities in the computation of anomalous dimensions in eqn. \eqref{cft op}.

\medskip
\noindent
As the method for obtaining the WGC bound from the binding energy of circular orbits is detailed in the previous subsection-(\ref{two}), we will be brief in what follows. 
The energy and angular momentum in the background eqn. (\ref{GBm}) can be obtained straightforwardly by evaluating the conditions $V(r_c) = 0  = V'(r_c)$ for circular orbits of radius $r_c$. However, the expressions are not very illuminating, and hence relegated to the Appendix. Also, our interest is in extracting the relevant pieces of the Energy and Angular momentum of probe particles in specific limits, from which the WGC bound could be seen. As in the previous subsection, we focus on the dynamics of charged test particles around a small black hole $(r_h \ll L)$  moving in a fairly large circular orbit, keeping in mind that the orbit radius is much smaller than the AdS-radius. The condition which describes this is, 
\begin{equation}
    \frac{M}{r^{d-3}} \ll \frac{r^2}{L^2} \ll 1 \, . \label{cond1}
\end{equation}
Due to the analytical complexity involved in evaluating the expressions, we prefer to work in a specific dimension, and chose $d=5$ (i.e., five dimensions) for now.  Restricting to the regime satisfying eqn. (\ref{cond1}), we extract the relevant terms from the expressions in appendix-(\ref{AppendixA}), and find the leading terms which contribute to energy and angular momentum to be,
\begin{equation}
    \hat E= 1+\left(\frac{3+2c_1}{3}\right)\frac{r^2}{L^2}\, ,
\end{equation}
\begin{equation}
    \hat l^2=\left(\frac{3+2c_1}{3}\right)\frac{r^4}{L^2}\left(1+\left(\frac{3}{3+2c_1}\right)\frac{L^2}{r^4}(M+\sqrt{3}\hat{q}Q(8c_2-1))\right)\, ,
\end{equation}
whereas in the large black hole regime $r_c\gg L, M^{\frac{1}{d-2}}$, we get 
\begin{equation}
    \hat E= 1+\left(\frac{3+2c_1}{3}\right)\frac{r^2}{L^2}+\frac{(3+2c_1)(2M+\sqrt{3}\hat{q}Q(8c_2-1))}{6L^2}
\end{equation}
\begin{equation}
     \hat l^2=\left(\frac{3+2c_1}{3}\right)\frac{r^4}{L^2}\left(1+\frac{2M+\sqrt{3}\hat{q}Q(8c_2-1)}{r^2}+\left(\frac{3}{3+2c_1}\right)\frac{L^2}{r^4}(M+\sqrt{3}\hat{q}Q(8c_2-1))\right)
\end{equation}
We note that taking $c_1,c_2 \to 0$ gives back the corresponding expressions of charged black holes in AdS (see eqns. (11) and (13) of ~\cite{Moitra:2023yyc}). Coordinate invariant expression of binding energy can now be obtained as before by finding $r_c$ as a perturbative expansion in terms of large $\hat l L_{eff}$ and yields,
\begin{equation}
    \hat E=1+\frac{\hat l}{L_{eff}}-\frac{1}{\hat l L_{eff}}\left(\frac{M}{2}-\frac{ \hat{q}Q(1-8c_2)}{2 k_d}\right)
\end{equation}
where, $L^2=L_{eff}^2\left(\frac{3+2c_1}{3}\right)$ and $k_d $ is a dimension dependent constant.
Now, using the map in eqn. (\ref{cft op})
we can get the anomalous dimension as,
\begin{equation}
    \gamma=-\Delta_q\left(\frac{M}{2}-\frac{ \hat{q}Q(1-8c_2)}{2 k_d}\right)\left(\frac{1}{L_{eff}}\right)^2\left(\frac{\Delta_q}{\mathcal{J}}\right)
\end{equation}
Demanding positivity of anomalous dimension gives
\begin{equation} \label{wgc2}
    \hat{q} \ge \frac{k_d}{(1-8c_2)}\frac{M_{ext}}{Q_{ext}} \, .
\end{equation}
Two comments are in order.
We first note that, the prefactor on the rhs of eqn. (\ref{wgc2}) expliclty depends only on $c_2$, and naively shows that the charge to mass ratio increases with the Wilson coefficients. However, this needs to be checked carefully, as the full set of Wilson coefficients are still present in the extremal mass $M_{ext}$, and extremal charge $Q_{ext}$. Since, we are working in the full higher derivative theory, the exact expressions for these extremal quantities cannot be derived analytically, and hence studying the dependence of $\hat q$ on the full set of Wilson coefficients, and comparison with earlier works~\cite{Cremonini:2019wdk}, is beyond the scope of the present work.  It would be nice to pursue this in future.  Second, although the analysis in this subsection was restricted to five space-time dimensions, we checked that the bound in eqn. (\ref{wgc2}) is valid in general dimensions, though the dimension dependent prefactor is different\footnote{For $d=4,5,6$, we find $k_d =  1/2, 1/\sqrt{3}, \sqrt{3/8}$, respectively.}.


\section{ISCO's and WGC} \label{three}

In this section, we numerically try to check the connection of ISCO and the WGC bound in Gauss-Bonnet black holes in AdS. A complete classification of regions around black hole space times where charged test particles can have possible circular motion is an interesting problem, which has been attempted both in asymptotically flat~\cite{Pugliese:2010ps,Pugliese:2011py,Pugliese:2013xfa} and AdS space-times~\cite{Chandrasekhar:2018sjg}. However, stable circular orbits do not always exist and there is a marginally stable orbit at a certain $r_{\rm isco}$, called the ISCO, which is the limiting radius. This radius can be found by imposing an additional condition 
\begin{equation} \label{iscoV}
V''(r_{\rm isco} ) = 0 \, .
\end{equation}
In asymptotically flat space times (RN or Gauss-Bonnet black holes), ISCOs exist only in four dimensions, and even here, they stop existing exactly when the charge to mass ratio is greater than unity \cite{Pugliese:2011py}. For neutral particles, ISCOs in a novel Gauss-Bonnet gravity in four dimensional asymptotically flat spacetime has been explored earlier in~\cite{Guo:2020zmf}. In the presence of a cosmological constant, one expects the situation to change. 
In~\cite{Moitra:2023yyc}, it was actually shown that in certain limits, it is possible to check for the existence of ISCOs in asymptotically AdS black holes, which have a connection with the WGC bound. This connection between ISCOs and WGC is quite interesting, especially due to the role played by AdS. In this section, we wish to check the effect of the Gauss-Bonnet coupling $\widetilde{\alpha}$ on the r$_{\rm isco}$ (the radius of ISCO) and its connection with the WGC bound. The actual equation to solve follows straightforwardly from eqn. (\ref{iscoV}), and it is given symbolically as:
\begin{equation} \label{vdp}
-2 A_t'' \hat q ( \hat E+ \hat q A_t )-2 A_t'^2
   \hat q^2-\frac{4 \hat l
   f'(r)}{r^3}+\left(\frac{\hat l}{r^2}+1\right)
   f''(r)+\frac{6 \hat l f(r)}{r^4} =0 \, ,
\end{equation}
Of course, before solving the above equation to find the value of $r_{\rm isco}$, one still has to insert the values of Energy and Angular momentum from eqns. (\ref{E1}) and (\ref{l1}) in the above eqn. (\ref{vdp}). 
\begin{figure}[h!]

	{\centering
		
		\subfloat[]{\includegraphics[width=2.8in]{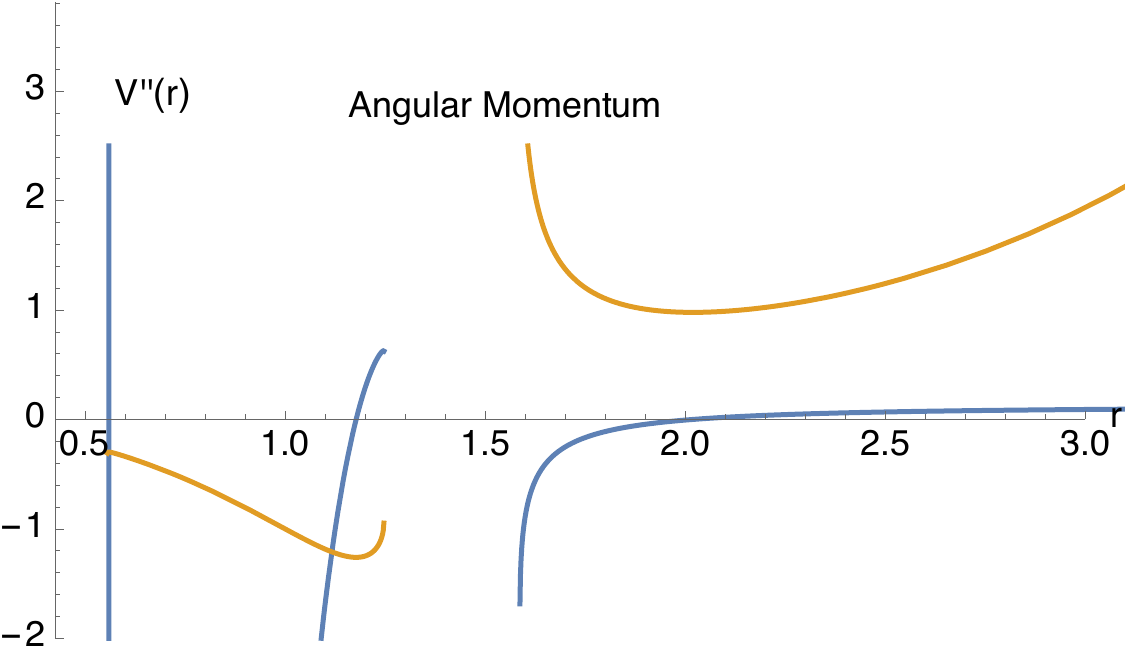}	\label{fig:isco}}\hspace{0.5cm}
		\subfloat[]{\includegraphics[width=2.8in]{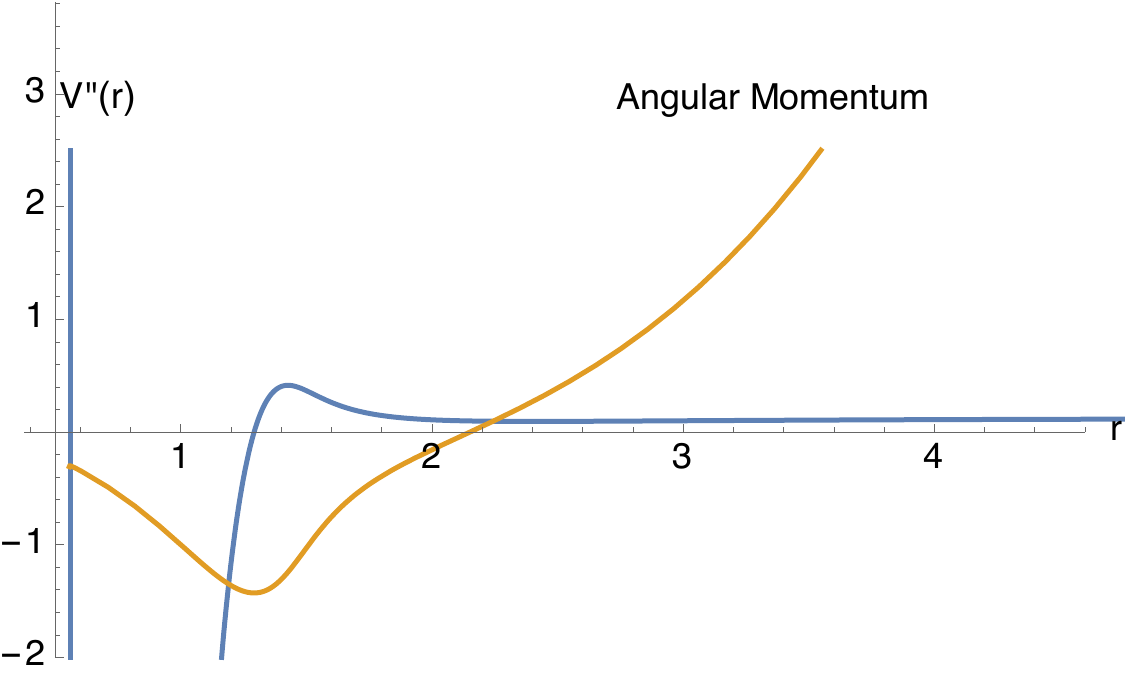}	\label{fig:noisco}}

		\caption{\footnotesize $V''(r)$ and angular momentum plotted w.r.t $r$ for various values of the charge to mass ratio $\hat q$. $V''(r)$ is evaluated together with $V(r)=V'(r)=0$, such that the zeros show the r$_{\rm isco}$ (the radius of ISCO).  For all the plots, $\widetilde{\alpha} = 0.01, L=10, d=5$ (AdS$_5$) (a) The charge to mass ratio $\hat q$ is below the WGC bound (eqn. (\ref{wgc})), resulting in one ISCO, for which angular momentum is positive. (b) The charge to mass ratio $\hat q$ is above the WGC bound (eqn. (\ref{wgc})). There are no ISCOs for which the angular momentum is positive}
	}	
\end{figure}
Though the resulting expression is not very illuminating to express here, some progress can be made numerically. The idea is to check for the existence of ISCOs, just above and below the bound obtained in eqn. (\ref{wgc}) for various values of the Gauss-Bonnet coupling parameter. First, by fixing $L$, we solve for the extremal charge in eqn. (\ref{Qext}) giving solutions for horizon radius. This can then be used in eqn. (\ref{7}) to fix the extremal mass, yielding an $\alpha$-dependent bound from eqn. (\ref{wgc}). We then considered various values of $\hat q$ below this bound, and checked for the zeros of eqn. (\ref{vdp}), keeping only those roots which result in positive angular momentum. The result of this analysis is shown in
figures (\ref{fig:isco}) and (\ref{fig:noisco}), which contains second derivatives of the potential $V''(r)$, and also angular momentum plotted w.r.t $r$, respectively, for values of the charge to mass ratio $q$, just below and above the WGC bound found in eqn. (\ref{wgc}). In plotting $V''(r)$, we made use of the conditions $V(r)=V'(r)=0$, such that the zeroes seen in the plots are r$_{\rm isco}$ (the radius of isco). However, as mentioned above, not all the zeroes are physical, and some can be eliminated by demanding the positivity of angular momentum. From figures (\ref{fig:isco}) and (\ref{fig:noisco}) we note that, the ISCOs stop existing precisely when the WGC bound is satisfied, confirming the earlier result~\cite{Moitra:2023yyc}, but now, in the presence of $\alpha$, which increases the bound as in eqn. (\ref{wgc}). This is satisfying to note and is valid for generic values of the Gauss-Bonnet coupling parameter $\alpha$. 
\begin{figure}[h!]
	
	{\centering
		
		{\includegraphics[width=3.0in]{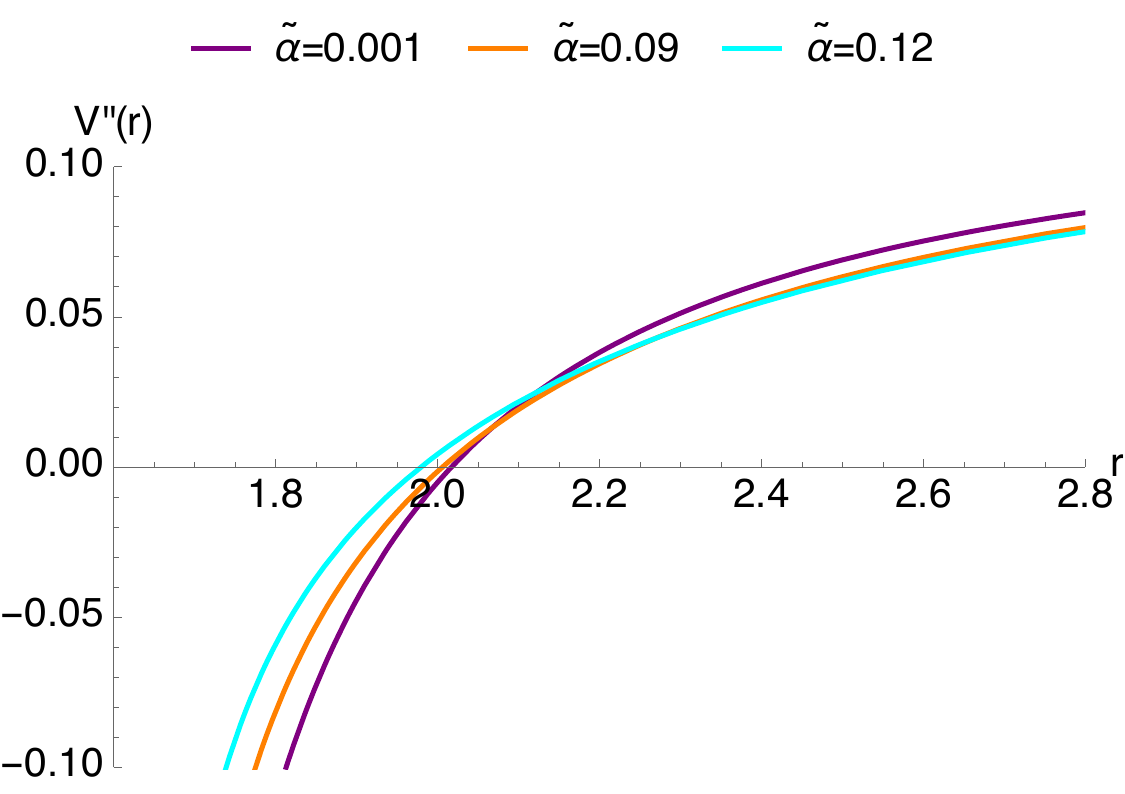}}	
		\caption{\footnotesize Plot of $V''(r)$ vs r (evaluated together with $V(r)=V'(r)=0$), the zeros of which show the r$_{\rm isco}$ (the radius of isco) for various values of the Gauss-Bonnet coupling parameter $\widetilde{\alpha}$. For all the plots, $ L=10, d=5$ (AdS$_5$) and $\hat q$ is taken to be below the WGC bound in eqn. (\ref{wgc}).}
		\label{fig:risco_alpha}	}
	
\end{figure}
Below the WGC bound, the corresponding radius r$_{\rm isco}$ decreases as $\alpha$ increases, as seen from figure-(\ref{fig:risco_alpha}). Momentarily, if one turns off the charges (neutral particle in an uncharged Gauss-Bonnet AdS black hole), the condition in eqn. (\ref{vdp}) for getting the ISCO simplifies a bit, and a perturbative calculation can be done, resulting in:
\begin{equation}
r_{\rm isco} = 2-\frac{33 \alpha }{100}-\frac{9 \alpha ^2}{25} \, - O[\alpha^3 ] \, .
\end{equation}
In obtaining the above, we assumed $d=5, M=1, L=4$, showing clearly a monotonous reduction in the ISCO radius. One can check that the leading value of $2$ in the above equation corresponds to radius of ISCO in the background of AdS Schwarzschild background~\cite{Berenstein:2020vlp}. The computations performed in this section, were for black holes of all sizes, and this reaffirms our faith in the bound in eqn. (\ref{wgc}) to be valid in general.  A similar analysis as done above, in the presence of more general higher derivative terms (from subsection-(\ref{three2})) is interesting, but is currently not tractable, particularly, because the input for this analysis requires exact expressions for extremal mass and charge in terms of the horizon radius  and the Wilson coefficients (in the full higher derivative theory). We leave this pursuit for future.

\section{Conclusions} \label{four}
In this article, we studied the connection between the innermost stable circular orbits (ISCOs) of charged particles in spherically symmetric AdS black holes and the Weak Gravity Conjecture (WGC) bound in higher derivative theories, and in particular, Gauss-Bonnet gravity in arbitrary dimensions, . Using the methods developed in~\cite{Berenstein:2020vlp,Dodelson:2022eiz,Moitra:2023yyc}, we studied a specific large radius limit well suited for capturing the effect of a finite, but small cosmological constant, and obtained the binding energy of charged probes in the limit important for small extremal black holes. It is known that significant simplifications occur in the conformal bootstrap analysis in the large spin limit~\cite{Alday:2007mf,Komargodski:2012ek,Alday:2015eya,Kulaxizi:2018dxo, Karlsson:2019qfi,Li:2019zba,Li:2020dqm}. In the current situation, the  anomalous dimensions in the  large spin (and charge) expansion can be obtained from studying the coordinate independent expressions for binding energy of probe particles~\cite{Berenstein:2020vlp,Moitra:2023yyc,Dodelson:2022eiz}.  Transcribing the results for binding energy into CFT quantities gives us the anomalous dimension of certain double twist operators with an effective AdS length scale corrected with higher derivative couplings. From the argument that the CFT anomalous dimensions of like charged operators are expected to be non-negative, results in a bound for charge to mass ratio of probes, which one identifies as the WGC bound as found in eqns. (\ref{wgc}), and (\ref{wgc2}) for Gauss-Bonnet theories and certain four derivative theories, respectively.   One understands that the CFT states giving rise to large charge to mass ratios in this case are expected to be narrow resonances and may not exactly be the states seen in the Hilbert space, though it would be interesting to study this further~\cite{Jafferis:2017zna,Alday:2015eya,Fitzpatrick:2014vua,Komargodski:2012ek}. Since, the ISCOs separate the stable from unstable orbits, it is interesting that they connect nicely with the WGC bound. The ISCOs stop existing precisely when the WGC bound is satisfied~\cite{Moitra:2023yyc}, which we find to be valid in the presence of Gauss-Bonnet coupling and should be checked in other general theories of gravity, with possible scalars etc.. The zero temperature limit gives extremal black holes, which contain the near horizon AdS$_2$ geometry. The AdS$_2$ Breitenlohner-Freedman bound can be violated by a charged scalar, but still satisfying the AdS$_{d}$ bound. It would be useful to check the scaling dimensions of bulk scalar fields in relation to the WGC bound~\cite{Breitenlohner:1982jf,Festuccia:2008zx,Hartnoll:2008vx,Gubser:2008px}. 

\medskip
\noindent
We compared our results with the computations done earlier in literature via two different methods. 
While our method is based on studying the charge to mass ratio of probe particles in AdS backgrounds, our results for the anomalous dimension in eqn. (\ref{ana}) and/or the WGC bound in eqn. (\ref{wgc}) are in agreement in appropriate limits with the recent computations for Schwarzschild AdS~\cite{Berenstein:2020vlp}, charged AdS~~\cite{Moitra:2023yyc} and neutral Gauss-Bonnet  black holes in AdS~\cite{Dodelson:2022eiz}. On the other hand, WGC bound has been studied earlier in higher derivative theories via a different method, with two important distinctions. First, the mass and charge of the extremal black hole in the two derivative theory are found exactly, and the contribution of higher derivative terms is added perturbatively to the two derivative result. Second, the computations are done for self-repulsive particles, especially in~\cite{Kats:2006xp,Cremonini:2009ih,Cremonini:2019wdk}. The effect of the Gauss-Bonnet coupling $\alpha$ (positive) and the Wilson coefficients, in general, is to increase the anomalous dimensions, as well as the charge to mass ratio, and ours bound in eqn. (\ref{wgc}) matches qualitatively with these expectations. However, in general, the series expansion of WGC bound we find in terms of higher derivative couplings, particularly from probe particles in Gauss-Bonnet AdS black hole backgrounds, does not match the corresponding expressions known for self-repulsive particles~\cite{Natsuume:1994hd,Kats:2006xp,Cano:2019oma,Cano:2019ycn,Cano:2021nzo,Ma:2021opb,Cheung:2018cwt,Hamada:2018dde,Bellazzini:2019xts,Charles:2019qqt,Jones:2019nev,Loges:2019jzs,Goon:2019faz,Cremonini:2019wdk,Chen:2020rov,Loges:2020trf,Bobev:2021oku,Arkani-Hamed:2021ajd,Cremonini:2021upd,Aalsma:2021qga,Amsel:2010aj}.  In the case of the bound in eqn. (\ref{wgc2}) for the four derivative theory, the exact expression we find cannot be expanded in a series, for comparison with earlier results~\cite{Cremonini:2019wdk}. These are interesting issues and it would be nice to pursue them in future.  

\medskip 
\noindent
We also checked that the radius of ISCOs decreases with the increase in the GB coupling parameter. It is of course important to obtain corrections to the WGC bound in Lovelock theories and in the presence of other higher derivative corrections in the electromagnetic  sector, as well as, generalizations to rotating black holes. Such studies might give a better understanding of black hole physics in the AdS/CFT context, as there are intricate connections between the WGC bound and microscopic physics. The connection between ISCOs and phase transitions of black holes in AdS is an interesting topic and it should be explored.

\section*{Acknowledgements}
We thank the editor for patience, and the anonymous referee for constructive suggestions which improved the draft.

\renewcommand{\thesection}{\Alph{section}}
\appendix
\section*{Appendices}

\section{Energy and angular momentum of probe particles in the charged AdS black holes in the four derivative theory } \label{AppendixA}
The expressions for energy and angular momentum (to be used in subsection-(\ref{three2})) in the background eqn. (\ref{GBm}) obtained by evaluating the conditions $V(r_c) = 0  = V'(r_c)$ for circular orbits of radius $r_c$ are given below, respectively:
\small
\begin{align*} 
    \hat{E} & = - \frac{\hat{q}Q(12r^6(8c_2-1)+L^2(-13Q^2c_1+48(Q^2-2Mr^2)c_2+72Q^2(2c_3+c_4)))}{8\sqrt{3}r^{8}}+ \\& \biggl(\biggl(1+\frac{L^2Q^2(\frac{26c_1}{2}+8c_2)}{r^6}\biggr)\biggl(1+\frac{Q^2}{4r^4}-\frac{M}{r^2}+\frac{r^2}{L^2}+\biggl(-\frac{191L^2Q^4}{96r^{14}}+\frac{31L^2MQ^2}{3r^{8}}+\frac{2L^2M^2}{r^6} \\ &-\frac{65Q^2}{3r^4}-\frac{16L^2Q^2}{r^4}+\frac{2r^2}{3L^2}\biggr)c_1+ \biggl(-\frac{9L^2Q^4}{4r^{10}}+\frac{10L^2MQ^2}{r^{8}}-\frac{12L^2Q^2}{r^6}-\frac{18Q^2}{r^4}\biggr)c_2- \\ & \frac{3L^2Q^4(2c_3+c_4)}{8r^{10}}\biggr)\biggl(-\hat{q}r\biggl(\frac{Q(-192L^2Mrc_2+72r^5(8c_2-1))}{8\sqrt{3}r^8}- \\ & \frac{Q(-12r^6(8c_2-1)+L^2(-13Q^2c_1+48(Q^2-2Mr^2)c_2+72Q^2(2c_3+c_4)))}{\sqrt{3}r^9}\biggr) + \\ & \surd \biggl(4\biggl(1+\frac{L^2Q^2(\frac{26c_1}{2}+8c_2)}{r^6}\biggr)\biggl(1+\frac{Q^2}{4r^4}-\frac{M}{r^2}+\frac{r^2}{L^2}+\biggl(-\frac{191L^2Q^4}{96r^{14}}+\frac{31L^2MQ^2}{3r^{8}}+\frac{2L^2M^2}{r^6} \\ &-\frac{65Q^2}{3r^4}-\frac{16L^2Q^2}{r^4}+\frac{2r^2}{3L^2}\biggr)c_1+ \biggl(-\frac{9L^2Q^4}{4r^{10}}+\frac{10L^2MQ^2}{r^{8}}-\frac{12L^2Q^2}{r^6}-\frac{18Q^2}{r^4}\biggr)c_2- \\ & \frac{3L^2Q^4(2c_3+c_4)}{8r^{10}}\biggr)-2r\biggl(\biggl(1+\frac{L^2Q^2(\frac{26c_1}{3}+8c_2)}{r^6}\biggr)\biggl(-\frac{Q^2}{r^5}+\frac{2M}{r^3}+\frac{2r}{L^2}+\biggl(\frac{955L^2Q^4}{48r^{11}}- \frac{248L^2MQ^2}{3r^9}- \\ & \frac{12L^2MQ^2}{r^7}+\frac{260Q^2}{3r^5}+\frac{64L^2Q^2}{r^5}+\frac{4r}{3L^2}\biggr)c_1 +\biggl(\frac{45L^2Q^4}{2r^{11}}-\frac{80L^2MQ^2}{r^9}+\frac{72L^2Q^2}{r^7}+\frac{72Q^2}{r^5}\biggr)c_2 + \\ & \frac{15L^2Q^4(2C_3+c_4)}{4r^{11}}\biggr) -\frac{1}{r^7}6L^2Q^2\biggl(\frac{26c_1}{3}+8c_2\biggr) \biggl(1+\frac{Q^2}{4r^4}-\frac{M}{r^2}+\frac{r^2}{L^2}+\biggl(-\frac{191L^2Q^4}{96r^{14}}+\frac{31L^2MQ^2}{3r^{8}} \\ & +\frac{2L^2M^2}{r^6} -\frac{65Q^2}{3r^4}-\frac{16L^2Q^2}{r^4}+\frac{2r^2}{3L^2}\biggr)c_1+ \biggl(-\frac{9L^2Q^4}{4r^{10}}+\frac{10L^2MQ^2}{r^{8}}-\frac{12L^2Q^2}{r^6}-\frac{18Q^2}{r^4}\biggr)c_2- \\ & \frac{3L^2Q^4(2c_3+c_4)}{8r^{10}}\biggr)\biggr)+\hat{q^2}r^2\biggl(\frac{Q(-192L^2Mrc_2+72r^5(8c_2-1))}{8\sqrt{3}r^8}- \\ & \frac{Q(12r^6(8c_2-1)+L^2(-13Q^2c_1+48(Q^2-2Mr^2)c_2)+72Q^2(2c_3+c_4))}{\sqrt{3}r^9}\biggr)^2\biggr)\biggr)\biggr) \Big /  \\ & \biggl(2\biggl(1+\frac{L^2Q^2(\frac{26c_1}{2}+8c_2)}{r^6}\biggr)\biggl(1+\frac{Q^2}{4r^4}-\frac{M}{r^2}+\frac{r^2}{L^2}+\biggl(-\frac{191L^2Q^4}{96r^{14}}+\frac{31L^2MQ^2}{3r^{8}}+\frac{2L^2M^2}{r^6} \\ &-\frac{65Q^2}{3r^4}-\frac{16L^2Q^2}{r^4}+\frac{2r^2}{3L^2}\biggr)c_1+ \biggl(-\frac{9L^2Q^4}{4r^{10}}+\frac{10L^2MQ^2}{r^{8}}-\frac{12L^2Q^2}{r^6}-\frac{18Q^2}{r^4}\biggr)c_2-  \frac{3L^2Q^4(2c_3+c_4)}{8r^{10}}\biggr) \\ & - r\biggl(\biggl(1+\frac{L^2Q^2(\frac{26c_1}{3}+8c_2)}{r^6}\biggr)\biggl(-\frac{Q^2}{r^5}+\frac{2M}{r^3}+\frac{2r}{L^2}+\biggl(\frac{955L^2Q^4}{48r^{11}}- \frac{248L^2MQ^2}{3r^9}- \\ & \frac{12L^2MQ^2}{r^7}+\frac{260Q^2}{3r^5}+\frac{64L^2Q^2}{r^5}+\frac{4r}{3L^2}\biggr)c_1 +\biggl(\frac{45L^2Q^4}{2r^{11}}-\frac{80L^2MQ^2}{r^9}+\frac{72L^2Q^2}{r^7}+\frac{72Q^2}{r^5}\biggr)c_2 + \\ & \frac{15L^2Q^4(2C_3+c_4)}{4r^{11}}\biggr) -\frac{1}{r^7}6L^2Q^2\biggl(\frac{26c_1}{3}+8c_2\biggr) \biggl(1+\frac{Q^2}{4r^4}-\frac{M}{r^2}+\frac{r^2}{L^2}+\biggl(-\frac{191L^2Q^4}{96r^{14}}+\frac{31L^2MQ^2}{3r^{8}} 
\end{align*}
\begin{align*}
    \\ & +\frac{2L^2M^2}{r^6} -\frac{65Q^2}{3r^4}-\frac{16L^2Q^2}{r^4}+\frac{2r^2}{3L^2}\biggr)c_1+ \biggl(-\frac{9L^2Q^4}{4r^{10}}+\frac{10L^2MQ^2}{r^{8}}-\frac{12L^2Q^2}{r^6}-\frac{18Q^2}{r^4}\biggr)c_2- \\ & \frac{3L^2Q^4(2c_3+c_4)}{8r^{10}}\biggr)\biggr)\biggr)
\end{align*}

 \begin{align*} 
     \hat{l}^2 &= \biggl(r^3\biggl(-r\biggl(\biggl(1+\frac{L^2Q^2(\frac{26c_1}{3}+8c_2)}{r^6}\biggr)\biggl(-\frac{Q^2}{r^5}+\frac{2M}{r^3}+\frac{2r}{L^2}+\biggl(\frac{955L^2Q^4}{48r^11}-\frac{248L^2MQ^2}{3r^9}-\frac{12L^2M^2}{r^7}+\\ & \frac{260Q^2}{3r^5}+ \frac{64L^2Q^2}{r^5}+\frac{4r}{3L^2}\biggr)c_1 +  \biggl(\frac{45L^2Q^4}{2r^{11}}-\frac{80L^2MQ^2}{r^9}+\frac{72L^2Q^2}{r^7}+\frac{72Q^2}{r^5}\biggr)c_2+\frac{15L^2Q^4(2c_3+c_4)}{4r^{11}}) - \\ &  \frac{1}{r^7}6L^2Q^2\biggl(\frac{26c_1}{3}+8c_2\biggr) \biggl( 1+\frac{Q^2}{4r^4}-\frac{M}{r^2}+\frac{r^2}{L^2}+\biggl(-\frac{191L^2Q^4}{96r^{10}}+\frac{31L^2MQ^2}{3r^8}+\frac{2L^2M^2}{r^6}-\frac{65Q^2}{3r^4}-\frac{16L^2Q^2}{r^4}+\\ &\frac{2r^2}{3L^2}\biggr)c_1  +\biggl( -\frac{9L^2Q^4}{4r^{10}}+\frac{10L^2MQ^2}{r^8}-\frac{12L^2Q^2}{r^6}-\frac{18Q^2}{r^4}\biggr)c_2-\frac{3L^2Q^4(2c_3+c_4)}{8r^{10}}\biggr)\biggr)^2 + 2\biggl(1+\frac{L^2Q^2(\frac{26c_1}{3}+8c_2)}{r^6}\biggr) \\ & \biggl(1+\frac{Q^2}{4r^4}-\frac{M}{r^2}+\frac{r^2}{L^2}+\biggl(-\frac{191L^2Q^4}{96r^{10}}+\frac{31L^2MQ^2}{3r^8}+\frac{2L^2M^2}{r^6}-\frac{65Q^2}{3r^4}-\frac{16L^2Q^2}{r^4}+\frac{2r^2}{3L^2}\biggr)c_1 + \\ & \biggl( -\frac{9L^2Q^4}{4r^{10}}+\frac{10L^2MQ^2}{r^8}-\frac{12L^2Q^2}{r^6}-\frac{18Q^2}{r^4}\biggr)c_2-\frac{3L^2Q^4(2c_3+c_4)}{8r^{10}}\biggr)\biggl(\biggl(1+\frac{L^2Q^2(\frac{26c_1}{3}+8c_2)}{r^6}\biggr) \\ & \biggl(-\frac{Q^2}{r^5}+\frac{2M}{r^3}+\frac{2r}{L^2}+\biggl(\frac{955L^2Q^4}{48r^11}-\frac{248L^2MQ^2}{3r^9}-\frac{12L^2M^2}{r^7}+\frac{260Q^2}{3r^5}+ \frac{64L^2Q^2}{r^5}+\frac{4r}{3L^2}\biggr)c_1 +\\ & \biggl(\frac{45L^2Q^4}{2r^{11}}-\frac{80L^2MQ^2}{r^9}+\frac{72L^2Q^2}{r^7}+\frac{72Q^2}{r^5}\biggr)c_2+\frac{15L^2Q^4(2c_3+c_4)}{4r^{11}}\biggr)-  \frac{1}{r^7} 6L^2Q^2\biggl(\frac{26c_1}{3}+8c_2\biggr) \\ & \biggl( 1+\frac{Q^2}{4r^4}-\frac{M}{r^2}+\frac{r^2}{L^2}+\biggl(-\frac{191L^2Q^4}{96r^{10}}+\frac{31L^2MQ^2}{3r^8}+\frac{2L^2M^2}{r^6}-\frac{65Q^2}{3r^4}-\frac{16L^2Q^2}{r^4}+\frac{2r^2}{3L^2}\biggr)c_1 +  \\ & \biggl( -\frac{9L^2Q^4}{4r^{10}}+\frac{10L^2MQ^2}{r^8}-\frac{12L^2Q^2}{r^6}-\frac{18Q^2}{r^4}\biggr)c_2 - \frac{3L^2Q^4(2c_3+c_4)}{8r^{10}}\biggr) + \hat{q}^2r\biggl(\frac{Q(-192L^2Mrc_2+72r^5(8c_2-1))}{8\sqrt{3}r^8} \\ & - \frac{Q(12r^6(8c_2-1)+L^2(-13Q^2c_1+48(Q^2-2Mr^2)c_2+72Q^2(2c_3+c_4)))}{\sqrt{3}r^9}\biggr)^2\biggr)- \\ & 2\hat{q} \biggr(\frac{Q(-192L^2Mrc_2+72r^5(8c_2-1))}{8\sqrt{3}r^8} -  \\ & \frac{Q(12r^6(8c_2-1)+L^2(-13Q^2c_1+48(Q^2-2Mr^2)c_2+72Q^2(2c_3+c_4)))}{\sqrt{3}r^9}\biggr) \surd \biggl(\biggl(1+\frac{L^2Q^2(\frac{26c_1}{3}+8c_2)}{r^6}\biggr)^2 
     \\ &  \biggl( 1+\frac{Q^2}{4r^4}-\frac{M}{r^2}+\frac{r^2}{L^2}+\biggl(-\frac{191L^2Q^4}{96r^{10}}+\frac{31L^2MQ^2}{3r^8}+\frac{2L^2M^2}{r^6}-\frac{65Q^2}{3r^4}-\frac{16L^2Q^2}{r^4}+\frac{2r^2}{3L^2}\biggr)c_1  \\ & \biggl( -\frac{9L^2Q^4}{4r^{10}}+\frac{10L^2MQ^2}{r^8}-\frac{12L^2Q^2}{r^6}-\frac{18Q^2}{r^4}\biggr)c_2 - \frac{3L^2Q^4(2c_3+c_4)}{8r^{10}}\biggr)^2 \biggl(4\biggl(1+\frac{L^2Q^2(\frac{26c_1}{3}+8c_2)}{r^6}\biggr) \\ & \biggl(1+\frac{Q^2}{4r^4}-\frac{M}{r^2}+\frac{r^2}{L^2}+\biggl(-\frac{191L^2Q^4}{96r^{10}}+\frac{31L^2MQ^2}{3r^8}+\frac{2L^2M^2}{r^6}-\frac{65Q^2}{3r^4}-\frac{16L^2Q^2}{r^4}+\frac{2r^2}{3L^2}\biggr)c_1 + \\ & \biggl( -\frac{9L^2Q^4}{4r^{10}}+\frac{10L^2MQ^2}{r^8}-\frac{12L^2Q^2}{r^6}-\frac{18Q^2}{r^4}\biggr)c_2-\frac{3L^2Q^4(2c_3+c_4)}{8r^{10}}\biggr) -2r \biggl(\biggl(1+\frac{L^2Q^2(\frac{26c_1}{3}+8c_2)}{r^6}\biggr) \\ & \biggl(-\frac{Q^2}{r^5}+\frac{2M}{r^3}+\frac{2r}{L^2}+\biggl(\frac{955L^2Q^4}{48r^11}-\frac{248L^2MQ^2}{3r^9}-\frac{12L^2M^2}{r^7}+\frac{260Q^2}{3r^5}+ \frac{64L^2Q^2}{r^5}+\frac{4r}{3L^2}\biggr)c_1 + \\ & \biggl(\frac{45L^2Q^4}{2r^{11}}-\frac{80L^2MQ^2}{r^9}+\frac{72L^2Q^2}{r^7}+\frac{72Q^2}{r^5}\biggr)c_2+\frac{15L^2Q^4(2c_3+c_4)}{4r^{11}}\biggr)-  \frac{1}{r^7} 6L^2Q^2\biggl(\frac{26c_1}{3}+8c_2\biggr) \\ & \biggl( 1+\frac{Q^2}{4r^4}-\frac{M}{r^2}+\frac{r^2}{L^2}+\biggl(-\frac{191L^2Q^4}{96r^{10}}+\frac{31L^2MQ^2}{3r^8}+\frac{2L^2M^2}{r^6}-\frac{65Q^2}{3r^4}-\frac{16L^2Q^2}{r^4}+\frac{2r^2}{3L^2}\biggr)c_1 + 
 \end{align*}
 
 \begin{align*}
\\ &  \biggl( -\frac{9L^2Q^4}{4r^{10}}+\frac{10L^2MQ^2}{r^8}-\frac{12L^2Q^2}{r^6}-\frac{18Q^2}{r^4}\biggr)c_2 - \frac{3L^2Q^4(2c_3+c_4)}{8r^{10}}\biggr) + \hat{q}^2r^2\biggl(\frac{Q(-192L^2Mrc_2+72r^5(8c_2-1))}{8\sqrt{3}r^8}  \\ & -\frac{Q(12r^6(8c_2-1)+L^2(-13Q^2c_1+48(Q^2-2Mr^2)c_2+72Q^2(2c_3+c_4)))}{\sqrt{3}r^9}\biggr)^2\biggr)\biggr)\biggr)\biggr) \Big /   \\ & \biggl(-2\biggl(1+\frac{L^2Q^2(\frac{26c_1}{3}+8c_2)}{r^6}\biggr) \biggl( 1+\frac{Q^2}{4r^4}-\frac{M}{r^2}+\frac{r^2}{L^2}+\biggl(-\frac{191L^2Q^4}{96r^{10}}+\frac{31L^2MQ^2}{3r^8}+\frac{2L^2M^2}{r^6}-\frac{65Q^2}{3r^4} \\ & -\frac{16L^2Q^2}{r^4}+\frac{2r^2}{3L^2}\biggr)c_1 +   \biggl( -\frac{9L^2Q^4}{4r^{10}}+\frac{10L^2MQ^2}{r^8}-\frac{12L^2Q^2}{r^6}-\frac{18Q^2}{r^4}\biggr)c_2 - \frac{3L^2Q^4(2c_3+c_4)}{8r^{10}}\biggr) \\ & + r \biggl(\biggl( 1+\frac{L^2Q^2(\frac{26c_1}{3}+8c_2)}{r^6}\biggr) \\ & \biggl(-\frac{Q^2}{r^5}+\frac{2M}{r^3}+\frac{2r}{L^2}+\biggl(\frac{955L^2Q^4}{48r^11}-\frac{248L^2MQ^2}{3r^9}-\frac{12L^2M^2}{r^7}+\frac{260Q^2}{3r^5}+ \frac{64L^2Q^2}{r^5}+\frac{4r}{3L^2}\biggr)c_1 +\\ & \biggl(\frac{45L^2Q^4}{2r^{11}}-\frac{80L^2MQ^2}{r^9}+\frac{72L^2Q^2}{r^7}+\frac{72Q^2}{r^5}\biggr)c_2+\frac{15L^2Q^4(2c_3+c_4)}{4r^{11}}\biggr)-  \frac{1}{r^7} 6L^2Q^2\biggl(\frac{26c_1}{3}+8c_2\biggr) \\ & \biggl( 1+\frac{Q^2}{4r^4}-\frac{M}{r^2}+\frac{r^2}{L^2}+\biggl(-\frac{191L^2Q^4}{96r^{10}}+\frac{31L^2MQ^2}{3r^8}+\frac{2L^2M^2}{r^6}-\frac{65Q^2}{3r^4}-\frac{16L^2Q^2}{r^4}+\frac{2r^2}{3L^2}\biggr)c_1 +  \\ & \biggl( -\frac{9L^2Q^4}{4r^{10}}+\frac{10L^2MQ^2}{r^8}-\frac{12L^2Q^2}{r^6}-\frac{18Q^2}{r^4}\biggr)c_2 - \frac{3L^2Q^4(2c_3+c_4)}{8r^{10}}\biggr)\biggr)\biggr)^2
 \end{align*}

\bibliographystyle{apsrev4-1}
\bibliography{wgc.bib}

\begin{thebibliography}{110}%
\makeatletter
\providecommand \@ifxundefined [1]{%
 \@ifx{#1\undefined}
}%
\providecommand \@ifnum [1]{%
 \ifnum #1\expandafter \@firstoftwo
 \else \expandafter \@secondoftwo
 \fi
}%
\providecommand \@ifx [1]{%
 \ifx #1\expandafter \@firstoftwo
 \else \expandafter \@secondoftwo
 \fi
}%
\providecommand \natexlab [1]{#1}%
\providecommand \enquote  [1]{``#1''}%
\providecommand \bibnamefont  [1]{#1}%
\providecommand \bibfnamefont [1]{#1}%
\providecommand \citenamefont [1]{#1}%
\providecommand \href@noop [0]{\@secondoftwo}%
\providecommand \href [0]{\begingroup \@sanitize@url \@href}%
\providecommand \@href[1]{\@@startlink{#1}\@@href}%
\providecommand \@@href[1]{\endgroup#1\@@endlink}%
\providecommand \@sanitize@url [0]{\catcode `\\12\catcode `\$12\catcode
  `\&12\catcode `\#12\catcode `\^12\catcode `\_12\catcode `\%12\relax}%
\providecommand \@@startlink[1]{}%
\providecommand \@@endlink[0]{}%
\providecommand \url  [0]{\begingroup\@sanitize@url \@url }%
\providecommand \@url [1]{\endgroup\@href {#1}{\urlprefix }}%
\providecommand \urlprefix  [0]{URL }%
\providecommand \Eprint [0]{\href }%
\providecommand \doibase [0]{http://dx.doi.org/}%
\providecommand \selectlanguage [0]{\@gobble}%
\providecommand \bibinfo  [0]{\@secondoftwo}%
\providecommand \bibfield  [0]{\@secondoftwo}%
\providecommand \translation [1]{[#1]}%
\providecommand \BibitemOpen [0]{}%
\providecommand \bibitemStop [0]{}%
\providecommand \bibitemNoStop [0]{.\EOS\space}%
\providecommand \EOS [0]{\spacefactor3000\relax}%
\providecommand \BibitemShut  [1]{\csname bibitem#1\endcsname}%
\let\auto@bib@innerbib\@empty
\bibitem [{\citenamefont {Berenstein}\ \emph {et~al.}(2021)\citenamefont
  {Berenstein}, \citenamefont {Li},\ and\ \citenamefont
  {Simon}}]{Berenstein:2020vlp}%
  \BibitemOpen
  \bibfield  {author} {\bibinfo {author} {\bibfnamefont {D.}~\bibnamefont
  {Berenstein}}, \bibinfo {author} {\bibfnamefont {Z.}~\bibnamefont {Li}}, \
  and\ \bibinfo {author} {\bibfnamefont {J.}~\bibnamefont {Simon}},\ }\href
  {\doibase 10.1088/1361-6382/abcaeb} {\bibfield  {journal} {\bibinfo
  {journal} {Class. Quant. Grav.}\ }\textbf {\bibinfo {volume} {38}},\ \bibinfo
  {pages} {045009} (\bibinfo {year} {2021})},\ \Eprint
  {http://arxiv.org/abs/2009.04500} {arXiv:2009.04500 [hep-th]} \BibitemShut
  {NoStop}%
\bibitem [{\citenamefont {Moitra}(2024)}]{Moitra:2023yyc}%
  \BibitemOpen
  \bibfield  {author} {\bibinfo {author} {\bibfnamefont {U.}~\bibnamefont
  {Moitra}},\ }\href {\doibase 10.1103/PhysRevD.109.L041903} {\bibfield
  {journal} {\bibinfo  {journal} {Phys. Rev. D}\ }\textbf {\bibinfo {volume}
  {109}},\ \bibinfo {pages} {L041903} (\bibinfo {year} {2024})},\ \Eprint
  {http://arxiv.org/abs/2305.08907} {arXiv:2305.08907 [hep-th]} \BibitemShut
  {NoStop}%
\bibitem [{\citenamefont {Dodelson}\ and\ \citenamefont
  {Zhiboedov}(2022)}]{Dodelson:2022eiz}%
  \BibitemOpen
  \bibfield  {author} {\bibinfo {author} {\bibfnamefont {M.}~\bibnamefont
  {Dodelson}}\ and\ \bibinfo {author} {\bibfnamefont {A.}~\bibnamefont
  {Zhiboedov}},\ }\href {\doibase 10.1007/JHEP12(2022)163} {\bibfield
  {journal} {\bibinfo  {journal} {JHEP}\ }\textbf {\bibinfo {volume} {12}},\
  \bibinfo {pages} {163} (\bibinfo {year} {2022})},\ \Eprint
  {http://arxiv.org/abs/2204.09749} {arXiv:2204.09749 [hep-th]} \BibitemShut
  {NoStop}%
\bibitem [{\citenamefont {Arkani-Hamed}\ \emph {et~al.}(2007)\citenamefont
  {Arkani-Hamed}, \citenamefont {Motl}, \citenamefont {Nicolis},\ and\
  \citenamefont {Vafa}}]{Arkani-Hamed:2006emk}%
  \BibitemOpen
  \bibfield  {author} {\bibinfo {author} {\bibfnamefont {N.}~\bibnamefont
  {Arkani-Hamed}}, \bibinfo {author} {\bibfnamefont {L.}~\bibnamefont {Motl}},
  \bibinfo {author} {\bibfnamefont {A.}~\bibnamefont {Nicolis}}, \ and\
  \bibinfo {author} {\bibfnamefont {C.}~\bibnamefont {Vafa}},\ }\href {\doibase
  10.1088/1126-6708/2007/06/060} {\bibfield  {journal} {\bibinfo  {journal}
  {JHEP}\ }\textbf {\bibinfo {volume} {06}},\ \bibinfo {pages} {060} (\bibinfo
  {year} {2007})},\ \Eprint {http://arxiv.org/abs/hep-th/0601001}
  {arXiv:hep-th/0601001} \BibitemShut {NoStop}%
\bibitem [{\citenamefont {Vafa}(2005)}]{Vafa:2005ui}%
  \BibitemOpen
  \bibfield  {author} {\bibinfo {author} {\bibfnamefont {C.}~\bibnamefont
  {Vafa}},\ }\href@noop {} {\  (\bibinfo {year} {2005})},\ \Eprint
  {http://arxiv.org/abs/hep-th/0509212} {arXiv:hep-th/0509212} \BibitemShut
  {NoStop}%
\bibitem [{\citenamefont {Brennan}\ \emph {et~al.}(2017)\citenamefont
  {Brennan}, \citenamefont {Carta},\ and\ \citenamefont
  {Vafa}}]{Brennan:2017rbf}%
  \BibitemOpen
  \bibfield  {author} {\bibinfo {author} {\bibfnamefont {T.~D.}\ \bibnamefont
  {Brennan}}, \bibinfo {author} {\bibfnamefont {F.}~\bibnamefont {Carta}}, \
  and\ \bibinfo {author} {\bibfnamefont {C.}~\bibnamefont {Vafa}},\ }\href
  {\doibase 10.22323/1.305.0015} {\bibfield  {journal} {\bibinfo  {journal}
  {PoS}\ }\textbf {\bibinfo {volume} {TASI2017}},\ \bibinfo {pages} {015}
  (\bibinfo {year} {2017})},\ \Eprint {http://arxiv.org/abs/1711.00864}
  {arXiv:1711.00864 [hep-th]} \BibitemShut {NoStop}%
\bibitem [{\citenamefont {Palti}(2019)}]{Palti:2019pca}%
  \BibitemOpen
  \bibfield  {author} {\bibinfo {author} {\bibfnamefont {E.}~\bibnamefont
  {Palti}},\ }\href {\doibase 10.1002/prop.201900037} {\bibfield  {journal}
  {\bibinfo  {journal} {Fortsch. Phys.}\ }\textbf {\bibinfo {volume} {67}},\
  \bibinfo {pages} {1900037} (\bibinfo {year} {2019})},\ \Eprint
  {http://arxiv.org/abs/1903.06239} {arXiv:1903.06239 [hep-th]} \BibitemShut
  {NoStop}%
\bibitem [{\citenamefont {van Beest}\ \emph {et~al.}(2022)\citenamefont {van
  Beest}, \citenamefont {Calder\'on-Infante}, \citenamefont {Mirfendereski},\
  and\ \citenamefont {Valenzuela}}]{vanBeest:2021lhn}%
  \BibitemOpen
  \bibfield  {author} {\bibinfo {author} {\bibfnamefont {M.}~\bibnamefont {van
  Beest}}, \bibinfo {author} {\bibfnamefont {J.}~\bibnamefont
  {Calder\'on-Infante}}, \bibinfo {author} {\bibfnamefont {D.}~\bibnamefont
  {Mirfendereski}}, \ and\ \bibinfo {author} {\bibfnamefont {I.}~\bibnamefont
  {Valenzuela}},\ }\href {\doibase 10.1016/j.physrep.2022.09.002} {\bibfield
  {journal} {\bibinfo  {journal} {Phys. Rept.}\ }\textbf {\bibinfo {volume}
  {989}},\ \bibinfo {pages} {1} (\bibinfo {year} {2022})},\ \Eprint
  {http://arxiv.org/abs/2102.01111} {arXiv:2102.01111 [hep-th]} \BibitemShut
  {NoStop}%
\bibitem [{\citenamefont {Harlow}\ \emph {et~al.}(2023)\citenamefont {Harlow},
  \citenamefont {Heidenreich}, \citenamefont {Reece},\ and\ \citenamefont
  {Rudelius}}]{Harlow:2022ich}%
  \BibitemOpen
  \bibfield  {author} {\bibinfo {author} {\bibfnamefont {D.}~\bibnamefont
  {Harlow}}, \bibinfo {author} {\bibfnamefont {B.}~\bibnamefont {Heidenreich}},
  \bibinfo {author} {\bibfnamefont {M.}~\bibnamefont {Reece}}, \ and\ \bibinfo
  {author} {\bibfnamefont {T.}~\bibnamefont {Rudelius}},\ }\href {\doibase
  10.1103/RevModPhys.95.035003} {\bibfield  {journal} {\bibinfo  {journal}
  {Rev. Mod. Phys.}\ }\textbf {\bibinfo {volume} {95}},\ \bibinfo {pages}
  {035003} (\bibinfo {year} {2023})},\ \Eprint
  {http://arxiv.org/abs/2201.08380} {arXiv:2201.08380 [hep-th]} \BibitemShut
  {NoStop}%
\bibitem [{\citenamefont {Natsuume}(1994)}]{Natsuume:1994hd}%
  \BibitemOpen
  \bibfield  {author} {\bibinfo {author} {\bibfnamefont {M.}~\bibnamefont
  {Natsuume}},\ }\href {\doibase 10.1103/PhysRevD.50.3949} {\bibfield
  {journal} {\bibinfo  {journal} {Phys. Rev. D}\ }\textbf {\bibinfo {volume}
  {50}},\ \bibinfo {pages} {3949} (\bibinfo {year} {1994})},\ \Eprint
  {http://arxiv.org/abs/hep-th/9406079} {arXiv:hep-th/9406079} \BibitemShut
  {NoStop}%
\bibitem [{\citenamefont {Kats}\ \emph {et~al.}(2007)\citenamefont {Kats},
  \citenamefont {Motl},\ and\ \citenamefont {Padi}}]{Kats:2006xp}%
  \BibitemOpen
  \bibfield  {author} {\bibinfo {author} {\bibfnamefont {Y.}~\bibnamefont
  {Kats}}, \bibinfo {author} {\bibfnamefont {L.}~\bibnamefont {Motl}}, \ and\
  \bibinfo {author} {\bibfnamefont {M.}~\bibnamefont {Padi}},\ }\href {\doibase
  10.1088/1126-6708/2007/12/068} {\bibfield  {journal} {\bibinfo  {journal}
  {JHEP}\ }\textbf {\bibinfo {volume} {12}},\ \bibinfo {pages} {068} (\bibinfo
  {year} {2007})},\ \Eprint {http://arxiv.org/abs/hep-th/0606100}
  {arXiv:hep-th/0606100} \BibitemShut {NoStop}%
\bibitem [{\citenamefont {Cano}\ \emph
  {et~al.}(2020{\natexlab{a}})\citenamefont {Cano}, \citenamefont
  {Ort\'\i{}n},\ and\ \citenamefont {Ramirez}}]{Cano:2019oma}%
  \BibitemOpen
  \bibfield  {author} {\bibinfo {author} {\bibfnamefont {P.~A.}\ \bibnamefont
  {Cano}}, \bibinfo {author} {\bibfnamefont {T.}~\bibnamefont {Ort\'\i{}n}}, \
  and\ \bibinfo {author} {\bibfnamefont {P.~F.}\ \bibnamefont {Ramirez}},\
  }\href {\doibase 10.1007/JHEP02(2020)175} {\bibfield  {journal} {\bibinfo
  {journal} {JHEP}\ }\textbf {\bibinfo {volume} {02}},\ \bibinfo {pages} {175}
  (\bibinfo {year} {2020}{\natexlab{a}})},\ \Eprint
  {http://arxiv.org/abs/1909.08530} {arXiv:1909.08530 [hep-th]} \BibitemShut
  {NoStop}%
\bibitem [{\citenamefont {Cano}\ \emph
  {et~al.}(2020{\natexlab{b}})\citenamefont {Cano}, \citenamefont {Chimento},
  \citenamefont {Linares}, \citenamefont {Ort\'\i{}n},\ and\ \citenamefont
  {Ram\'\i{}rez}}]{Cano:2019ycn}%
  \BibitemOpen
  \bibfield  {author} {\bibinfo {author} {\bibfnamefont {P.~A.}\ \bibnamefont
  {Cano}}, \bibinfo {author} {\bibfnamefont {S.}~\bibnamefont {Chimento}},
  \bibinfo {author} {\bibfnamefont {R.}~\bibnamefont {Linares}}, \bibinfo
  {author} {\bibfnamefont {T.}~\bibnamefont {Ort\'\i{}n}}, \ and\ \bibinfo
  {author} {\bibfnamefont {P.~F.}\ \bibnamefont {Ram\'\i{}rez}},\ }\href
  {\doibase 10.1007/JHEP02(2020)031} {\bibfield  {journal} {\bibinfo  {journal}
  {JHEP}\ }\textbf {\bibinfo {volume} {02}},\ \bibinfo {pages} {031} (\bibinfo
  {year} {2020}{\natexlab{b}})},\ \Eprint {http://arxiv.org/abs/1910.14324}
  {arXiv:1910.14324 [hep-th]} \BibitemShut {NoStop}%
\bibitem [{\citenamefont {Cano}\ \emph {et~al.}(2022)\citenamefont {Cano},
  \citenamefont {Ort\'\i{}n}, \citenamefont {Ruip\'erez},\ and\ \citenamefont
  {Zatti}}]{Cano:2021nzo}%
  \BibitemOpen
  \bibfield  {author} {\bibinfo {author} {\bibfnamefont {P.~A.}\ \bibnamefont
  {Cano}}, \bibinfo {author} {\bibfnamefont {T.}~\bibnamefont {Ort\'\i{}n}},
  \bibinfo {author} {\bibfnamefont {A.}~\bibnamefont {Ruip\'erez}}, \ and\
  \bibinfo {author} {\bibfnamefont {M.}~\bibnamefont {Zatti}},\ }\href
  {\doibase 10.1007/JHEP03(2022)103} {\bibfield  {journal} {\bibinfo  {journal}
  {JHEP}\ }\textbf {\bibinfo {volume} {03}},\ \bibinfo {pages} {103} (\bibinfo
  {year} {2022})},\ \Eprint {http://arxiv.org/abs/2111.15579} {arXiv:2111.15579
  [hep-th]} \BibitemShut {NoStop}%
\bibitem [{\citenamefont {Ma}\ \emph {et~al.}(2022)\citenamefont {Ma},
  \citenamefont {Pang},\ and\ \citenamefont {L\"u}}]{Ma:2021opb}%
  \BibitemOpen
  \bibfield  {author} {\bibinfo {author} {\bibfnamefont {L.}~\bibnamefont
  {Ma}}, \bibinfo {author} {\bibfnamefont {Y.}~\bibnamefont {Pang}}, \ and\
  \bibinfo {author} {\bibfnamefont {H.}~\bibnamefont {L\"u}},\ }\href {\doibase
  10.1007/JHEP01(2022)157} {\bibfield  {journal} {\bibinfo  {journal} {JHEP}\
  }\textbf {\bibinfo {volume} {01}},\ \bibinfo {pages} {157} (\bibinfo {year}
  {2022})},\ \Eprint {http://arxiv.org/abs/2110.03129} {arXiv:2110.03129
  [hep-th]} \BibitemShut {NoStop}%
\bibitem [{\citenamefont {Cheung}\ \emph {et~al.}(2018)\citenamefont {Cheung},
  \citenamefont {Liu},\ and\ \citenamefont {Remmen}}]{Cheung:2018cwt}%
  \BibitemOpen
  \bibfield  {author} {\bibinfo {author} {\bibfnamefont {C.}~\bibnamefont
  {Cheung}}, \bibinfo {author} {\bibfnamefont {J.}~\bibnamefont {Liu}}, \ and\
  \bibinfo {author} {\bibfnamefont {G.~N.}\ \bibnamefont {Remmen}},\ }\href
  {\doibase 10.1007/JHEP10(2018)004} {\bibfield  {journal} {\bibinfo  {journal}
  {JHEP}\ }\textbf {\bibinfo {volume} {10}},\ \bibinfo {pages} {004} (\bibinfo
  {year} {2018})},\ \Eprint {http://arxiv.org/abs/1801.08546} {arXiv:1801.08546
  [hep-th]} \BibitemShut {NoStop}%
\bibitem [{\citenamefont {Hamada}\ \emph {et~al.}(2019)\citenamefont {Hamada},
  \citenamefont {Noumi},\ and\ \citenamefont {Shiu}}]{Hamada:2018dde}%
  \BibitemOpen
  \bibfield  {author} {\bibinfo {author} {\bibfnamefont {Y.}~\bibnamefont
  {Hamada}}, \bibinfo {author} {\bibfnamefont {T.}~\bibnamefont {Noumi}}, \
  and\ \bibinfo {author} {\bibfnamefont {G.}~\bibnamefont {Shiu}},\ }\href
  {\doibase 10.1103/PhysRevLett.123.051601} {\bibfield  {journal} {\bibinfo
  {journal} {Phys. Rev. Lett.}\ }\textbf {\bibinfo {volume} {123}},\ \bibinfo
  {pages} {051601} (\bibinfo {year} {2019})},\ \Eprint
  {http://arxiv.org/abs/1810.03637} {arXiv:1810.03637 [hep-th]} \BibitemShut
  {NoStop}%
\bibitem [{\citenamefont {Bellazzini}\ \emph {et~al.}(2019)\citenamefont
  {Bellazzini}, \citenamefont {Lewandowski},\ and\ \citenamefont
  {Serra}}]{Bellazzini:2019xts}%
  \BibitemOpen
  \bibfield  {author} {\bibinfo {author} {\bibfnamefont {B.}~\bibnamefont
  {Bellazzini}}, \bibinfo {author} {\bibfnamefont {M.}~\bibnamefont
  {Lewandowski}}, \ and\ \bibinfo {author} {\bibfnamefont {J.}~\bibnamefont
  {Serra}},\ }\href {\doibase 10.1103/PhysRevLett.123.251103} {\bibfield
  {journal} {\bibinfo  {journal} {Phys. Rev. Lett.}\ }\textbf {\bibinfo
  {volume} {123}},\ \bibinfo {pages} {251103} (\bibinfo {year} {2019})},\
  \Eprint {http://arxiv.org/abs/1902.03250} {arXiv:1902.03250 [hep-th]}
  \BibitemShut {NoStop}%
\bibitem [{\citenamefont {Charles}(2019)}]{Charles:2019qqt}%
  \BibitemOpen
  \bibfield  {author} {\bibinfo {author} {\bibfnamefont {A.~M.}\ \bibnamefont
  {Charles}},\ }\href@noop {} {\  (\bibinfo {year} {2019})},\ \Eprint
  {http://arxiv.org/abs/1906.07734} {arXiv:1906.07734 [hep-th]} \BibitemShut
  {NoStop}%
\bibitem [{\citenamefont {Jones}\ and\ \citenamefont
  {McPeak}(2020)}]{Jones:2019nev}%
  \BibitemOpen
  \bibfield  {author} {\bibinfo {author} {\bibfnamefont {C.~R.~T.}\
  \bibnamefont {Jones}}\ and\ \bibinfo {author} {\bibfnamefont
  {B.}~\bibnamefont {McPeak}},\ }\href {\doibase 10.1007/JHEP06(2020)140}
  {\bibfield  {journal} {\bibinfo  {journal} {JHEP}\ }\textbf {\bibinfo
  {volume} {06}},\ \bibinfo {pages} {140} (\bibinfo {year} {2020})},\ \Eprint
  {http://arxiv.org/abs/1908.10452} {arXiv:1908.10452 [hep-th]} \BibitemShut
  {NoStop}%
\bibitem [{\citenamefont {Loges}\ \emph
  {et~al.}(2020{\natexlab{a}})\citenamefont {Loges}, \citenamefont {Noumi},\
  and\ \citenamefont {Shiu}}]{Loges:2019jzs}%
  \BibitemOpen
  \bibfield  {author} {\bibinfo {author} {\bibfnamefont {G.~J.}\ \bibnamefont
  {Loges}}, \bibinfo {author} {\bibfnamefont {T.}~\bibnamefont {Noumi}}, \ and\
  \bibinfo {author} {\bibfnamefont {G.}~\bibnamefont {Shiu}},\ }\href {\doibase
  10.1103/PhysRevD.102.046010} {\bibfield  {journal} {\bibinfo  {journal}
  {Phys. Rev. D}\ }\textbf {\bibinfo {volume} {102}},\ \bibinfo {pages}
  {046010} (\bibinfo {year} {2020}{\natexlab{a}})},\ \Eprint
  {http://arxiv.org/abs/1909.01352} {arXiv:1909.01352 [hep-th]} \BibitemShut
  {NoStop}%
\bibitem [{\citenamefont {Goon}\ and\ \citenamefont
  {Penco}(2020)}]{Goon:2019faz}%
  \BibitemOpen
  \bibfield  {author} {\bibinfo {author} {\bibfnamefont {G.}~\bibnamefont
  {Goon}}\ and\ \bibinfo {author} {\bibfnamefont {R.}~\bibnamefont {Penco}},\
  }\href {\doibase 10.1103/PhysRevLett.124.101103} {\bibfield  {journal}
  {\bibinfo  {journal} {Phys. Rev. Lett.}\ }\textbf {\bibinfo {volume} {124}},\
  \bibinfo {pages} {101103} (\bibinfo {year} {2020})},\ \Eprint
  {http://arxiv.org/abs/1909.05254} {arXiv:1909.05254 [hep-th]} \BibitemShut
  {NoStop}%
\bibitem [{\citenamefont {Cremonini}\ \emph {et~al.}(2020)\citenamefont
  {Cremonini}, \citenamefont {Jones}, \citenamefont {Liu},\ and\ \citenamefont
  {McPeak}}]{Cremonini:2019wdk}%
  \BibitemOpen
  \bibfield  {author} {\bibinfo {author} {\bibfnamefont {S.}~\bibnamefont
  {Cremonini}}, \bibinfo {author} {\bibfnamefont {C.~R.~T.}\ \bibnamefont
  {Jones}}, \bibinfo {author} {\bibfnamefont {J.~T.}\ \bibnamefont {Liu}}, \
  and\ \bibinfo {author} {\bibfnamefont {B.}~\bibnamefont {McPeak}},\ }\href
  {\doibase 10.1007/JHEP09(2020)003} {\bibfield  {journal} {\bibinfo  {journal}
  {JHEP}\ }\textbf {\bibinfo {volume} {09}},\ \bibinfo {pages} {003} (\bibinfo
  {year} {2020})},\ \Eprint {http://arxiv.org/abs/1912.11161} {arXiv:1912.11161
  [hep-th]} \BibitemShut {NoStop}%
\bibitem [{\citenamefont {Chen}\ \emph {et~al.}(2020)\citenamefont {Chen},
  \citenamefont {Hong},\ and\ \citenamefont {Tao}}]{Chen:2020rov}%
  \BibitemOpen
  \bibfield  {author} {\bibinfo {author} {\bibfnamefont {Q.}~\bibnamefont
  {Chen}}, \bibinfo {author} {\bibfnamefont {W.}~\bibnamefont {Hong}}, \ and\
  \bibinfo {author} {\bibfnamefont {J.}~\bibnamefont {Tao}},\ }\href@noop {} {\
   (\bibinfo {year} {2020})},\ \Eprint {http://arxiv.org/abs/2005.00747}
  {arXiv:2005.00747 [gr-qc]} \BibitemShut {NoStop}%
\bibitem [{\citenamefont {Loges}\ \emph
  {et~al.}(2020{\natexlab{b}})\citenamefont {Loges}, \citenamefont {Noumi},\
  and\ \citenamefont {Shiu}}]{Loges:2020trf}%
  \BibitemOpen
  \bibfield  {author} {\bibinfo {author} {\bibfnamefont {G.~J.}\ \bibnamefont
  {Loges}}, \bibinfo {author} {\bibfnamefont {T.}~\bibnamefont {Noumi}}, \ and\
  \bibinfo {author} {\bibfnamefont {G.}~\bibnamefont {Shiu}},\ }\href {\doibase
  10.1007/JHEP11(2020)008} {\bibfield  {journal} {\bibinfo  {journal} {JHEP}\
  }\textbf {\bibinfo {volume} {11}},\ \bibinfo {pages} {008} (\bibinfo {year}
  {2020}{\natexlab{b}})},\ \Eprint {http://arxiv.org/abs/2006.06696}
  {arXiv:2006.06696 [hep-th]} \BibitemShut {NoStop}%
\bibitem [{\citenamefont {Bobev}\ \emph {et~al.}(2021)\citenamefont {Bobev},
  \citenamefont {Charles}, \citenamefont {Hristov},\ and\ \citenamefont
  {Reys}}]{Bobev:2021oku}%
  \BibitemOpen
  \bibfield  {author} {\bibinfo {author} {\bibfnamefont {N.}~\bibnamefont
  {Bobev}}, \bibinfo {author} {\bibfnamefont {A.~M.}\ \bibnamefont {Charles}},
  \bibinfo {author} {\bibfnamefont {K.}~\bibnamefont {Hristov}}, \ and\
  \bibinfo {author} {\bibfnamefont {V.}~\bibnamefont {Reys}},\ }\href {\doibase
  10.1007/JHEP08(2021)173} {\bibfield  {journal} {\bibinfo  {journal} {JHEP}\
  }\textbf {\bibinfo {volume} {08}},\ \bibinfo {pages} {173} (\bibinfo {year}
  {2021})},\ \Eprint {http://arxiv.org/abs/2106.04581} {arXiv:2106.04581
  [hep-th]} \BibitemShut {NoStop}%
\bibitem [{\citenamefont {Arkani-Hamed}\ \emph {et~al.}(2022)\citenamefont
  {Arkani-Hamed}, \citenamefont {Huang}, \citenamefont {Liu},\ and\
  \citenamefont {Remmen}}]{Arkani-Hamed:2021ajd}%
  \BibitemOpen
  \bibfield  {author} {\bibinfo {author} {\bibfnamefont {N.}~\bibnamefont
  {Arkani-Hamed}}, \bibinfo {author} {\bibfnamefont {Y.-t.}\ \bibnamefont
  {Huang}}, \bibinfo {author} {\bibfnamefont {J.-Y.}\ \bibnamefont {Liu}}, \
  and\ \bibinfo {author} {\bibfnamefont {G.~N.}\ \bibnamefont {Remmen}},\
  }\href {\doibase 10.1007/JHEP03(2022)083} {\bibfield  {journal} {\bibinfo
  {journal} {JHEP}\ }\textbf {\bibinfo {volume} {03}},\ \bibinfo {pages} {083}
  (\bibinfo {year} {2022})},\ \Eprint {http://arxiv.org/abs/2109.13937}
  {arXiv:2109.13937 [hep-th]} \BibitemShut {NoStop}%
\bibitem [{\citenamefont {Cremonini}\ \emph {et~al.}(2022)\citenamefont
  {Cremonini}, \citenamefont {Jones}, \citenamefont {Liu}, \citenamefont
  {McPeak},\ and\ \citenamefont {Tang}}]{Cremonini:2021upd}%
  \BibitemOpen
  \bibfield  {author} {\bibinfo {author} {\bibfnamefont {S.}~\bibnamefont
  {Cremonini}}, \bibinfo {author} {\bibfnamefont {C.~R.~T.}\ \bibnamefont
  {Jones}}, \bibinfo {author} {\bibfnamefont {J.~T.}\ \bibnamefont {Liu}},
  \bibinfo {author} {\bibfnamefont {B.}~\bibnamefont {McPeak}}, \ and\ \bibinfo
  {author} {\bibfnamefont {Y.}~\bibnamefont {Tang}},\ }\href {\doibase
  10.1007/JHEP03(2022)013} {\bibfield  {journal} {\bibinfo  {journal} {JHEP}\
  }\textbf {\bibinfo {volume} {03}},\ \bibinfo {pages} {013} (\bibinfo {year}
  {2022})},\ \Eprint {http://arxiv.org/abs/2110.10178} {arXiv:2110.10178
  [hep-th]} \BibitemShut {NoStop}%
\bibitem [{\citenamefont {Aalsma}(2022)}]{Aalsma:2021qga}%
  \BibitemOpen
  \bibfield  {author} {\bibinfo {author} {\bibfnamefont {L.}~\bibnamefont
  {Aalsma}},\ }\href {\doibase 10.1103/PhysRevD.105.066022} {\bibfield
  {journal} {\bibinfo  {journal} {Phys. Rev. D}\ }\textbf {\bibinfo {volume}
  {105}},\ \bibinfo {pages} {066022} (\bibinfo {year} {2022})},\ \Eprint
  {http://arxiv.org/abs/2111.04201} {arXiv:2111.04201 [hep-th]} \BibitemShut
  {NoStop}%
\bibitem [{\citenamefont {Amsel}\ and\ \citenamefont
  {Gorbonos}(2010)}]{Amsel:2010aj}%
  \BibitemOpen
  \bibfield  {author} {\bibinfo {author} {\bibfnamefont {A.~J.}\ \bibnamefont
  {Amsel}}\ and\ \bibinfo {author} {\bibfnamefont {D.}~\bibnamefont
  {Gorbonos}},\ }\href {\doibase 10.1007/JHEP11(2010)033} {\bibfield  {journal}
  {\bibinfo  {journal} {JHEP}\ }\textbf {\bibinfo {volume} {11}},\ \bibinfo
  {pages} {033} (\bibinfo {year} {2010})},\ \Eprint
  {http://arxiv.org/abs/1005.4718} {arXiv:1005.4718 [hep-th]} \BibitemShut
  {NoStop}%
\bibitem [{\citenamefont {Bowick}\ \emph {et~al.}(1986)\citenamefont {Bowick},
  \citenamefont {Smolin},\ and\ \citenamefont {Wijewardhana}}]{Bowick:1985af}%
  \BibitemOpen
  \bibfield  {author} {\bibinfo {author} {\bibfnamefont {M.~J.}\ \bibnamefont
  {Bowick}}, \bibinfo {author} {\bibfnamefont {L.}~\bibnamefont {Smolin}}, \
  and\ \bibinfo {author} {\bibfnamefont {L.~C.~R.}\ \bibnamefont
  {Wijewardhana}},\ }\href {\doibase 10.1103/PhysRevLett.56.424} {\bibfield
  {journal} {\bibinfo  {journal} {Phys. Rev. Lett.}\ }\textbf {\bibinfo
  {volume} {56}},\ \bibinfo {pages} {424} (\bibinfo {year} {1986})}\BibitemShut
  {NoStop}%
\bibitem [{\citenamefont {Susskind}(1993)}]{Susskind:1993ws}%
  \BibitemOpen
  \bibfield  {author} {\bibinfo {author} {\bibfnamefont {L.}~\bibnamefont
  {Susskind}},\ }\href@noop {} {\ ,\ \bibinfo {pages} {118} (\bibinfo {year}
  {1993})},\ \Eprint {http://arxiv.org/abs/hep-th/9309145}
  {arXiv:hep-th/9309145} \BibitemShut {NoStop}%
\bibitem [{\citenamefont {Horowitz}\ and\ \citenamefont
  {Polchinski}(1997)}]{Horowitz:1996nw}%
  \BibitemOpen
  \bibfield  {author} {\bibinfo {author} {\bibfnamefont {G.~T.}\ \bibnamefont
  {Horowitz}}\ and\ \bibinfo {author} {\bibfnamefont {J.}~\bibnamefont
  {Polchinski}},\ }\href {\doibase 10.1103/PhysRevD.55.6189} {\bibfield
  {journal} {\bibinfo  {journal} {Phys. Rev. D}\ }\textbf {\bibinfo {volume}
  {55}},\ \bibinfo {pages} {6189} (\bibinfo {year} {1997})},\ \Eprint
  {http://arxiv.org/abs/hep-th/9612146} {arXiv:hep-th/9612146} \BibitemShut
  {NoStop}%
\bibitem [{\citenamefont {Heidenreich}\ \emph {et~al.}(2017)\citenamefont
  {Heidenreich}, \citenamefont {Reece},\ and\ \citenamefont
  {Rudelius}}]{Heidenreich:2016aqi}%
  \BibitemOpen
  \bibfield  {author} {\bibinfo {author} {\bibfnamefont {B.}~\bibnamefont
  {Heidenreich}}, \bibinfo {author} {\bibfnamefont {M.}~\bibnamefont {Reece}},
  \ and\ \bibinfo {author} {\bibfnamefont {T.}~\bibnamefont {Rudelius}},\
  }\href {\doibase 10.1007/JHEP08(2017)025} {\bibfield  {journal} {\bibinfo
  {journal} {JHEP}\ }\textbf {\bibinfo {volume} {08}},\ \bibinfo {pages} {025}
  (\bibinfo {year} {2017})},\ \Eprint {http://arxiv.org/abs/1606.08437}
  {arXiv:1606.08437 [hep-th]} \BibitemShut {NoStop}%
\bibitem [{\citenamefont {Montero}\ \emph {et~al.}(2016)\citenamefont
  {Montero}, \citenamefont {Shiu},\ and\ \citenamefont
  {Soler}}]{Montero:2016tif}%
  \BibitemOpen
  \bibfield  {author} {\bibinfo {author} {\bibfnamefont {M.}~\bibnamefont
  {Montero}}, \bibinfo {author} {\bibfnamefont {G.}~\bibnamefont {Shiu}}, \
  and\ \bibinfo {author} {\bibfnamefont {P.}~\bibnamefont {Soler}},\ }\href
  {\doibase 10.1007/JHEP10(2016)159} {\bibfield  {journal} {\bibinfo  {journal}
  {JHEP}\ }\textbf {\bibinfo {volume} {10}},\ \bibinfo {pages} {159} (\bibinfo
  {year} {2016})},\ \Eprint {http://arxiv.org/abs/1606.08438} {arXiv:1606.08438
  [hep-th]} \BibitemShut {NoStop}%
\bibitem [{\citenamefont {Lee}\ \emph {et~al.}(2018)\citenamefont {Lee},
  \citenamefont {Lerche},\ and\ \citenamefont {Weigand}}]{Lee:2018urn}%
  \BibitemOpen
  \bibfield  {author} {\bibinfo {author} {\bibfnamefont {S.-J.}\ \bibnamefont
  {Lee}}, \bibinfo {author} {\bibfnamefont {W.}~\bibnamefont {Lerche}}, \ and\
  \bibinfo {author} {\bibfnamefont {T.}~\bibnamefont {Weigand}},\ }\href
  {\doibase 10.1007/JHEP10(2018)164} {\bibfield  {journal} {\bibinfo  {journal}
  {JHEP}\ }\textbf {\bibinfo {volume} {10}},\ \bibinfo {pages} {164} (\bibinfo
  {year} {2018})},\ \Eprint {http://arxiv.org/abs/1808.05958} {arXiv:1808.05958
  [hep-th]} \BibitemShut {NoStop}%
\bibitem [{\citenamefont {Lee}\ \emph {et~al.}(2019)\citenamefont {Lee},
  \citenamefont {Lerche},\ and\ \citenamefont {Weigand}}]{Lee:2019tst}%
  \BibitemOpen
  \bibfield  {author} {\bibinfo {author} {\bibfnamefont {S.-J.}\ \bibnamefont
  {Lee}}, \bibinfo {author} {\bibfnamefont {W.}~\bibnamefont {Lerche}}, \ and\
  \bibinfo {author} {\bibfnamefont {T.}~\bibnamefont {Weigand}},\ }\href
  {\doibase 10.1007/JHEP08(2019)104} {\bibfield  {journal} {\bibinfo  {journal}
  {JHEP}\ }\textbf {\bibinfo {volume} {08}},\ \bibinfo {pages} {104} (\bibinfo
  {year} {2019})},\ \Eprint {http://arxiv.org/abs/1901.08065} {arXiv:1901.08065
  [hep-th]} \BibitemShut {NoStop}%
\bibitem [{\citenamefont {Aalsma}\ \emph {et~al.}(2019)\citenamefont {Aalsma},
  \citenamefont {Cole},\ and\ \citenamefont {Shiu}}]{Aalsma:2019ryi}%
  \BibitemOpen
  \bibfield  {author} {\bibinfo {author} {\bibfnamefont {L.}~\bibnamefont
  {Aalsma}}, \bibinfo {author} {\bibfnamefont {A.}~\bibnamefont {Cole}}, \ and\
  \bibinfo {author} {\bibfnamefont {G.}~\bibnamefont {Shiu}},\ }\href {\doibase
  10.1007/JHEP08(2019)022} {\bibfield  {journal} {\bibinfo  {journal} {JHEP}\
  }\textbf {\bibinfo {volume} {08}},\ \bibinfo {pages} {022} (\bibinfo {year}
  {2019})},\ \Eprint {http://arxiv.org/abs/1905.06956} {arXiv:1905.06956
  [hep-th]} \BibitemShut {NoStop}%
\bibitem [{\citenamefont {Klaewer}\ \emph {et~al.}(2021)\citenamefont
  {Klaewer}, \citenamefont {Lee}, \citenamefont {Weigand},\ and\ \citenamefont
  {Wiesner}}]{Klaewer:2020lfg}%
  \BibitemOpen
  \bibfield  {author} {\bibinfo {author} {\bibfnamefont {D.}~\bibnamefont
  {Klaewer}}, \bibinfo {author} {\bibfnamefont {S.-J.}\ \bibnamefont {Lee}},
  \bibinfo {author} {\bibfnamefont {T.}~\bibnamefont {Weigand}}, \ and\
  \bibinfo {author} {\bibfnamefont {M.}~\bibnamefont {Wiesner}},\ }\href
  {\doibase 10.1007/JHEP03(2021)252} {\bibfield  {journal} {\bibinfo  {journal}
  {JHEP}\ }\textbf {\bibinfo {volume} {03}},\ \bibinfo {pages} {252} (\bibinfo
  {year} {2021})},\ \Eprint {http://arxiv.org/abs/2011.00024} {arXiv:2011.00024
  [hep-th]} \BibitemShut {NoStop}%
\bibitem [{\citenamefont {Cheung}\ and\ \citenamefont
  {Remmen}(2014)}]{Cheung:2014ega}%
  \BibitemOpen
  \bibfield  {author} {\bibinfo {author} {\bibfnamefont {C.}~\bibnamefont
  {Cheung}}\ and\ \bibinfo {author} {\bibfnamefont {G.~N.}\ \bibnamefont
  {Remmen}},\ }\href {\doibase 10.1007/JHEP12(2014)087} {\bibfield  {journal}
  {\bibinfo  {journal} {JHEP}\ }\textbf {\bibinfo {volume} {12}},\ \bibinfo
  {pages} {087} (\bibinfo {year} {2014})},\ \Eprint
  {http://arxiv.org/abs/1407.7865} {arXiv:1407.7865 [hep-th]} \BibitemShut
  {NoStop}%
\bibitem [{\citenamefont {Andriolo}\ \emph {et~al.}(2018)\citenamefont
  {Andriolo}, \citenamefont {Junghans}, \citenamefont {Noumi},\ and\
  \citenamefont {Shiu}}]{Andriolo:2018lvp}%
  \BibitemOpen
  \bibfield  {author} {\bibinfo {author} {\bibfnamefont {S.}~\bibnamefont
  {Andriolo}}, \bibinfo {author} {\bibfnamefont {D.}~\bibnamefont {Junghans}},
  \bibinfo {author} {\bibfnamefont {T.}~\bibnamefont {Noumi}}, \ and\ \bibinfo
  {author} {\bibfnamefont {G.}~\bibnamefont {Shiu}},\ }\href {\doibase
  10.1002/prop.201800020} {\bibfield  {journal} {\bibinfo  {journal} {Fortsch.
  Phys.}\ }\textbf {\bibinfo {volume} {66}},\ \bibinfo {pages} {1800020}
  (\bibinfo {year} {2018})},\ \Eprint {http://arxiv.org/abs/1802.04287}
  {arXiv:1802.04287 [hep-th]} \BibitemShut {NoStop}%
\bibitem [{\citenamefont {Chen}\ \emph {et~al.}(2019)\citenamefont {Chen},
  \citenamefont {Huang}, \citenamefont {Noumi},\ and\ \citenamefont
  {Wen}}]{Chen:2019qvr}%
  \BibitemOpen
  \bibfield  {author} {\bibinfo {author} {\bibfnamefont {W.-M.}\ \bibnamefont
  {Chen}}, \bibinfo {author} {\bibfnamefont {Y.-T.}\ \bibnamefont {Huang}},
  \bibinfo {author} {\bibfnamefont {T.}~\bibnamefont {Noumi}}, \ and\ \bibinfo
  {author} {\bibfnamefont {C.}~\bibnamefont {Wen}},\ }\href {\doibase
  10.1103/PhysRevD.100.025016} {\bibfield  {journal} {\bibinfo  {journal}
  {Phys. Rev. D}\ }\textbf {\bibinfo {volume} {100}},\ \bibinfo {pages}
  {025016} (\bibinfo {year} {2019})},\ \Eprint
  {http://arxiv.org/abs/1901.11480} {arXiv:1901.11480 [hep-th]} \BibitemShut
  {NoStop}%
\bibitem [{\citenamefont {Alberte}\ \emph {et~al.}(2021)\citenamefont
  {Alberte}, \citenamefont {de~Rham}, \citenamefont {Jaitly},\ and\
  \citenamefont {Tolley}}]{Alberte:2020bdz}%
  \BibitemOpen
  \bibfield  {author} {\bibinfo {author} {\bibfnamefont {L.}~\bibnamefont
  {Alberte}}, \bibinfo {author} {\bibfnamefont {C.}~\bibnamefont {de~Rham}},
  \bibinfo {author} {\bibfnamefont {S.}~\bibnamefont {Jaitly}}, \ and\ \bibinfo
  {author} {\bibfnamefont {A.~J.}\ \bibnamefont {Tolley}},\ }\href {\doibase
  10.1103/PhysRevD.103.125020} {\bibfield  {journal} {\bibinfo  {journal}
  {Phys. Rev. D}\ }\textbf {\bibinfo {volume} {103}},\ \bibinfo {pages}
  {125020} (\bibinfo {year} {2021})},\ \Eprint
  {http://arxiv.org/abs/2012.05798} {arXiv:2012.05798 [hep-th]} \BibitemShut
  {NoStop}%
\bibitem [{\citenamefont {Aoki}\ \emph {et~al.}(2021)\citenamefont {Aoki},
  \citenamefont {Loc}, \citenamefont {Noumi},\ and\ \citenamefont
  {Tokuda}}]{Aoki:2021ckh}%
  \BibitemOpen
  \bibfield  {author} {\bibinfo {author} {\bibfnamefont {K.}~\bibnamefont
  {Aoki}}, \bibinfo {author} {\bibfnamefont {T.~Q.}\ \bibnamefont {Loc}},
  \bibinfo {author} {\bibfnamefont {T.}~\bibnamefont {Noumi}}, \ and\ \bibinfo
  {author} {\bibfnamefont {J.}~\bibnamefont {Tokuda}},\ }\href {\doibase
  10.1103/PhysRevLett.127.091602} {\bibfield  {journal} {\bibinfo  {journal}
  {Phys. Rev. Lett.}\ }\textbf {\bibinfo {volume} {127}},\ \bibinfo {pages}
  {091602} (\bibinfo {year} {2021})},\ \Eprint
  {http://arxiv.org/abs/2104.09682} {arXiv:2104.09682 [hep-th]} \BibitemShut
  {NoStop}%
\bibitem [{\citenamefont {Noumi}\ and\ \citenamefont
  {Tokuda}(2021)}]{Noumi:2021uuv}%
  \BibitemOpen
  \bibfield  {author} {\bibinfo {author} {\bibfnamefont {T.}~\bibnamefont
  {Noumi}}\ and\ \bibinfo {author} {\bibfnamefont {J.}~\bibnamefont {Tokuda}},\
  }\href {\doibase 10.1103/PhysRevD.104.066022} {\bibfield  {journal} {\bibinfo
   {journal} {Phys. Rev. D}\ }\textbf {\bibinfo {volume} {104}},\ \bibinfo
  {pages} {066022} (\bibinfo {year} {2021})},\ \Eprint
  {http://arxiv.org/abs/2105.01436} {arXiv:2105.01436 [hep-th]} \BibitemShut
  {NoStop}%
\bibitem [{\citenamefont {Noumi}\ \emph {et~al.}(2023)\citenamefont {Noumi},
  \citenamefont {Sato},\ and\ \citenamefont {Tokuda}}]{Noumi:2022zht}%
  \BibitemOpen
  \bibfield  {author} {\bibinfo {author} {\bibfnamefont {T.}~\bibnamefont
  {Noumi}}, \bibinfo {author} {\bibfnamefont {S.}~\bibnamefont {Sato}}, \ and\
  \bibinfo {author} {\bibfnamefont {J.}~\bibnamefont {Tokuda}},\ }\href
  {\doibase 10.1103/PhysRevD.108.056013} {\bibfield  {journal} {\bibinfo
  {journal} {Phys. Rev. D}\ }\textbf {\bibinfo {volume} {108}},\ \bibinfo
  {pages} {056013} (\bibinfo {year} {2023})},\ \Eprint
  {http://arxiv.org/abs/2205.12835} {arXiv:2205.12835 [hep-th]} \BibitemShut
  {NoStop}%
\bibitem [{\citenamefont {Heidenreich}\ \emph {et~al.}(2016)\citenamefont
  {Heidenreich}, \citenamefont {Reece},\ and\ \citenamefont
  {Rudelius}}]{Heidenreich:2015nta}%
  \BibitemOpen
  \bibfield  {author} {\bibinfo {author} {\bibfnamefont {B.}~\bibnamefont
  {Heidenreich}}, \bibinfo {author} {\bibfnamefont {M.}~\bibnamefont {Reece}},
  \ and\ \bibinfo {author} {\bibfnamefont {T.}~\bibnamefont {Rudelius}},\
  }\href {\doibase 10.1007/JHEP02(2016)140} {\bibfield  {journal} {\bibinfo
  {journal} {JHEP}\ }\textbf {\bibinfo {volume} {02}},\ \bibinfo {pages} {140}
  (\bibinfo {year} {2016})},\ \Eprint {http://arxiv.org/abs/1509.06374}
  {arXiv:1509.06374 [hep-th]} \BibitemShut {NoStop}%
\bibitem [{\citenamefont {Maldacena}(1998)}]{Maldacena:1997re}%
  \BibitemOpen
  \bibfield  {author} {\bibinfo {author} {\bibfnamefont {J.~M.}\ \bibnamefont
  {Maldacena}},\ }\href {\doibase 10.4310/ATMP.1998.v2.n2.a1} {\bibfield
  {journal} {\bibinfo  {journal} {Adv. Theor. Math. Phys.}\ }\textbf {\bibinfo
  {volume} {2}},\ \bibinfo {pages} {231} (\bibinfo {year} {1998})},\ \Eprint
  {http://arxiv.org/abs/hep-th/9711200} {arXiv:hep-th/9711200} \BibitemShut
  {NoStop}%
\bibitem [{\citenamefont {Gubser}\ \emph {et~al.}(1998)\citenamefont {Gubser},
  \citenamefont {Klebanov},\ and\ \citenamefont {Polyakov}}]{Gubser:1998bc}%
  \BibitemOpen
  \bibfield  {author} {\bibinfo {author} {\bibfnamefont {S.~S.}\ \bibnamefont
  {Gubser}}, \bibinfo {author} {\bibfnamefont {I.~R.}\ \bibnamefont
  {Klebanov}}, \ and\ \bibinfo {author} {\bibfnamefont {A.~M.}\ \bibnamefont
  {Polyakov}},\ }\href {\doibase 10.1016/S0370-2693(98)00377-3} {\bibfield
  {journal} {\bibinfo  {journal} {Phys. Lett. B}\ }\textbf {\bibinfo {volume}
  {428}},\ \bibinfo {pages} {105} (\bibinfo {year} {1998})},\ \Eprint
  {http://arxiv.org/abs/hep-th/9802109} {arXiv:hep-th/9802109} \BibitemShut
  {NoStop}%
\bibitem [{\citenamefont {Witten}(1998)}]{Witten:1998qj}%
  \BibitemOpen
  \bibfield  {author} {\bibinfo {author} {\bibfnamefont {E.}~\bibnamefont
  {Witten}},\ }\href {\doibase 10.4310/ATMP.1998.v2.n2.a2} {\bibfield
  {journal} {\bibinfo  {journal} {Adv. Theor. Math. Phys.}\ }\textbf {\bibinfo
  {volume} {2}},\ \bibinfo {pages} {253} (\bibinfo {year} {1998})},\ \Eprint
  {http://arxiv.org/abs/hep-th/9802150} {arXiv:hep-th/9802150} \BibitemShut
  {NoStop}%
\bibitem [{\citenamefont {Fidkowski}\ \emph {et~al.}(2004)\citenamefont
  {Fidkowski}, \citenamefont {Hubeny}, \citenamefont {Kleban},\ and\
  \citenamefont {Shenker}}]{Fidkowski_2004}%
  \BibitemOpen
  \bibfield  {author} {\bibinfo {author} {\bibfnamefont {L.}~\bibnamefont
  {Fidkowski}}, \bibinfo {author} {\bibfnamefont {V.}~\bibnamefont {Hubeny}},
  \bibinfo {author} {\bibfnamefont {M.}~\bibnamefont {Kleban}}, \ and\ \bibinfo
  {author} {\bibfnamefont {S.}~\bibnamefont {Shenker}},\ }\href {\doibase
  10.1088/1126-6708/2004/02/014} {\bibfield  {journal} {\bibinfo  {journal}
  {Journal of High Energy Physics}\ }\textbf {\bibinfo {volume} {2004}},\
  \bibinfo {pages} {014–014} (\bibinfo {year} {2004})}\BibitemShut {NoStop}%
\bibitem [{\citenamefont {Cardoso}\ \emph {et~al.}(2009)\citenamefont
  {Cardoso}, \citenamefont {Miranda}, \citenamefont {Berti}, \citenamefont
  {Witek},\ and\ \citenamefont {Zanchin}}]{PhysRevD.79.064016}%
  \BibitemOpen
  \bibfield  {author} {\bibinfo {author} {\bibfnamefont {V.}~\bibnamefont
  {Cardoso}}, \bibinfo {author} {\bibfnamefont {A.~S.}\ \bibnamefont
  {Miranda}}, \bibinfo {author} {\bibfnamefont {E.}~\bibnamefont {Berti}},
  \bibinfo {author} {\bibfnamefont {H.}~\bibnamefont {Witek}}, \ and\ \bibinfo
  {author} {\bibfnamefont {V.~T.}\ \bibnamefont {Zanchin}},\ }\href {\doibase
  10.1103/PhysRevD.79.064016} {\bibfield  {journal} {\bibinfo  {journal} {Phys.
  Rev. D}\ }\textbf {\bibinfo {volume} {79}},\ \bibinfo {pages} {064016}
  (\bibinfo {year} {2009})}\BibitemShut {NoStop}%
\bibitem [{\citenamefont {Kinoshita}\ \emph {et~al.}(2023)\citenamefont
  {Kinoshita}, \citenamefont {Murata},\ and\ \citenamefont
  {Takeda}}]{Kinoshita:2023hgc}%
  \BibitemOpen
  \bibfield  {author} {\bibinfo {author} {\bibfnamefont {S.}~\bibnamefont
  {Kinoshita}}, \bibinfo {author} {\bibfnamefont {K.}~\bibnamefont {Murata}}, \
  and\ \bibinfo {author} {\bibfnamefont {D.}~\bibnamefont {Takeda}},\ }\href
  {\doibase 10.1007/JHEP10(2023)074} {\bibfield  {journal} {\bibinfo  {journal}
  {JHEP}\ }\textbf {\bibinfo {volume} {10}},\ \bibinfo {pages} {074} (\bibinfo
  {year} {2023})},\ \Eprint {http://arxiv.org/abs/2304.01936} {arXiv:2304.01936
  [hep-th]} \BibitemShut {NoStop}%
\bibitem [{\citenamefont {\v{C}eplak}\ \emph {et~al.}(2024)\citenamefont
  {\v{C}eplak}, \citenamefont {Liu}, \citenamefont {Parnachev},\ and\
  \citenamefont {Valach}}]{Ceplak:2024bja}%
  \BibitemOpen
  \bibfield  {author} {\bibinfo {author} {\bibfnamefont {N.}~\bibnamefont
  {\v{C}eplak}}, \bibinfo {author} {\bibfnamefont {H.}~\bibnamefont {Liu}},
  \bibinfo {author} {\bibfnamefont {A.}~\bibnamefont {Parnachev}}, \ and\
  \bibinfo {author} {\bibfnamefont {S.}~\bibnamefont {Valach}},\ }\href
  {\doibase 10.1007/JHEP10(2024)105} {\bibfield  {journal} {\bibinfo  {journal}
  {JHEP}\ }\textbf {\bibinfo {volume} {10}},\ \bibinfo {pages} {105} (\bibinfo
  {year} {2024})},\ \Eprint {http://arxiv.org/abs/2404.17286} {arXiv:2404.17286
  [hep-th]} \BibitemShut {NoStop}%
\bibitem [{\citenamefont {Hashimoto}\ \emph {et~al.}(2023)\citenamefont
  {Hashimoto}, \citenamefont {Sugiura}, \citenamefont {Sugiyama},\ and\
  \citenamefont {Yoda}}]{Hashimoto:2023buz}%
  \BibitemOpen
  \bibfield  {author} {\bibinfo {author} {\bibfnamefont {K.}~\bibnamefont
  {Hashimoto}}, \bibinfo {author} {\bibfnamefont {K.}~\bibnamefont {Sugiura}},
  \bibinfo {author} {\bibfnamefont {K.}~\bibnamefont {Sugiyama}}, \ and\
  \bibinfo {author} {\bibfnamefont {T.}~\bibnamefont {Yoda}},\ }\href {\doibase
  10.1007/JHEP10(2023)149} {\bibfield  {journal} {\bibinfo  {journal} {JHEP}\
  }\textbf {\bibinfo {volume} {10}},\ \bibinfo {pages} {149} (\bibinfo {year}
  {2023})},\ \Eprint {http://arxiv.org/abs/2307.00237} {arXiv:2307.00237
  [hep-th]} \BibitemShut {NoStop}%
\bibitem [{\citenamefont {Sekino}\ and\ \citenamefont
  {Susskind}(2008)}]{Sekino:2008he}%
  \BibitemOpen
  \bibfield  {author} {\bibinfo {author} {\bibfnamefont {Y.}~\bibnamefont
  {Sekino}}\ and\ \bibinfo {author} {\bibfnamefont {L.}~\bibnamefont
  {Susskind}},\ }\href {\doibase 10.1088/1126-6708/2008/10/065} {\bibfield
  {journal} {\bibinfo  {journal} {JHEP}\ }\textbf {\bibinfo {volume} {10}},\
  \bibinfo {pages} {065} (\bibinfo {year} {2008})},\ \Eprint
  {http://arxiv.org/abs/0808.2096} {arXiv:0808.2096 [hep-th]} \BibitemShut
  {NoStop}%
\bibitem [{\citenamefont {Abajo-Arrastia}\ \emph {et~al.}(2010)\citenamefont
  {Abajo-Arrastia}, \citenamefont {Aparicio},\ and\ \citenamefont
  {Lopez}}]{Abajo-Arrastia:2010ajo}%
  \BibitemOpen
  \bibfield  {author} {\bibinfo {author} {\bibfnamefont {J.}~\bibnamefont
  {Abajo-Arrastia}}, \bibinfo {author} {\bibfnamefont {J.}~\bibnamefont
  {Aparicio}}, \ and\ \bibinfo {author} {\bibfnamefont {E.}~\bibnamefont
  {Lopez}},\ }\href {\doibase 10.1007/JHEP11(2010)149} {\bibfield  {journal}
  {\bibinfo  {journal} {JHEP}\ }\textbf {\bibinfo {volume} {11}},\ \bibinfo
  {pages} {149} (\bibinfo {year} {2010})},\ \Eprint
  {http://arxiv.org/abs/1006.4090} {arXiv:1006.4090 [hep-th]} \BibitemShut
  {NoStop}%
\bibitem [{\citenamefont {Balasubramanian}\ \emph {et~al.}(2011)\citenamefont
  {Balasubramanian}, \citenamefont {Bernamonti}, \citenamefont {de~Boer},
  \citenamefont {Copland}, \citenamefont {Craps}, \citenamefont
  {Keski-Vakkuri}, \citenamefont {Muller}, \citenamefont {Schafer},
  \citenamefont {Shigemori},\ and\ \citenamefont
  {Staessens}}]{Balasubramanian:2011ur}%
  \BibitemOpen
  \bibfield  {author} {\bibinfo {author} {\bibfnamefont {V.}~\bibnamefont
  {Balasubramanian}}, \bibinfo {author} {\bibfnamefont {A.}~\bibnamefont
  {Bernamonti}}, \bibinfo {author} {\bibfnamefont {J.}~\bibnamefont {de~Boer}},
  \bibinfo {author} {\bibfnamefont {N.}~\bibnamefont {Copland}}, \bibinfo
  {author} {\bibfnamefont {B.}~\bibnamefont {Craps}}, \bibinfo {author}
  {\bibfnamefont {E.}~\bibnamefont {Keski-Vakkuri}}, \bibinfo {author}
  {\bibfnamefont {B.}~\bibnamefont {Muller}}, \bibinfo {author} {\bibfnamefont
  {A.}~\bibnamefont {Schafer}}, \bibinfo {author} {\bibfnamefont
  {M.}~\bibnamefont {Shigemori}}, \ and\ \bibinfo {author} {\bibfnamefont
  {W.}~\bibnamefont {Staessens}},\ }\href {\doibase 10.1103/PhysRevD.84.026010}
  {\bibfield  {journal} {\bibinfo  {journal} {Phys. Rev. D}\ }\textbf {\bibinfo
  {volume} {84}},\ \bibinfo {pages} {026010} (\bibinfo {year} {2011})},\
  \Eprint {http://arxiv.org/abs/1103.2683} {arXiv:1103.2683 [hep-th]}
  \BibitemShut {NoStop}%
\bibitem [{\citenamefont {Nozaki}\ \emph {et~al.}(2013)\citenamefont {Nozaki},
  \citenamefont {Numasawa},\ and\ \citenamefont {Takayanagi}}]{Nozaki:2013wia}%
  \BibitemOpen
  \bibfield  {author} {\bibinfo {author} {\bibfnamefont {M.}~\bibnamefont
  {Nozaki}}, \bibinfo {author} {\bibfnamefont {T.}~\bibnamefont {Numasawa}}, \
  and\ \bibinfo {author} {\bibfnamefont {T.}~\bibnamefont {Takayanagi}},\
  }\href {\doibase 10.1007/JHEP05(2013)080} {\bibfield  {journal} {\bibinfo
  {journal} {JHEP}\ }\textbf {\bibinfo {volume} {05}},\ \bibinfo {pages} {080}
  (\bibinfo {year} {2013})},\ \Eprint {http://arxiv.org/abs/1302.5703}
  {arXiv:1302.5703 [hep-th]} \BibitemShut {NoStop}%
\bibitem [{\citenamefont {Shenker}\ and\ \citenamefont
  {Stanford}(2014)}]{Shenker:2013pqa}%
  \BibitemOpen
  \bibfield  {author} {\bibinfo {author} {\bibfnamefont {S.~H.}\ \bibnamefont
  {Shenker}}\ and\ \bibinfo {author} {\bibfnamefont {D.}~\bibnamefont
  {Stanford}},\ }\href {\doibase 10.1007/JHEP03(2014)067} {\bibfield  {journal}
  {\bibinfo  {journal} {JHEP}\ }\textbf {\bibinfo {volume} {03}},\ \bibinfo
  {pages} {067} (\bibinfo {year} {2014})},\ \Eprint
  {http://arxiv.org/abs/1306.0622} {arXiv:1306.0622 [hep-th]} \BibitemShut
  {NoStop}%
\bibitem [{\citenamefont {Festuccia}\ and\ \citenamefont
  {Liu}(2009)}]{Festuccia:2008zx}%
  \BibitemOpen
  \bibfield  {author} {\bibinfo {author} {\bibfnamefont {G.}~\bibnamefont
  {Festuccia}}\ and\ \bibinfo {author} {\bibfnamefont {H.}~\bibnamefont
  {Liu}},\ }\href {\doibase 10.1166/asl.2009.1029} {\bibfield  {journal}
  {\bibinfo  {journal} {Adv. Sci. Lett.}\ }\textbf {\bibinfo {volume} {2}},\
  \bibinfo {pages} {221} (\bibinfo {year} {2009})},\ \Eprint
  {http://arxiv.org/abs/0811.1033} {arXiv:0811.1033 [gr-qc]} \BibitemShut
  {NoStop}%
\bibitem [{\citenamefont {Fitzpatrick}\ \emph {et~al.}(2014)\citenamefont
  {Fitzpatrick}, \citenamefont {Kaplan},\ and\ \citenamefont
  {Walters}}]{Fitzpatrick:2014vua}%
  \BibitemOpen
  \bibfield  {author} {\bibinfo {author} {\bibfnamefont {A.~L.}\ \bibnamefont
  {Fitzpatrick}}, \bibinfo {author} {\bibfnamefont {J.}~\bibnamefont {Kaplan}},
  \ and\ \bibinfo {author} {\bibfnamefont {M.~T.}\ \bibnamefont {Walters}},\
  }\href {\doibase 10.1007/JHEP08(2014)145} {\bibfield  {journal} {\bibinfo
  {journal} {JHEP}\ }\textbf {\bibinfo {volume} {08}},\ \bibinfo {pages} {145}
  (\bibinfo {year} {2014})},\ \Eprint {http://arxiv.org/abs/1403.6829}
  {arXiv:1403.6829 [hep-th]} \BibitemShut {NoStop}%
\bibitem [{\citenamefont {Cruz}\ \emph {et~al.}(1994)\citenamefont {Cruz},
  \citenamefont {Martinez},\ and\ \citenamefont {Pena}}]{Cruz:1994ir}%
  \BibitemOpen
  \bibfield  {author} {\bibinfo {author} {\bibfnamefont {N.}~\bibnamefont
  {Cruz}}, \bibinfo {author} {\bibfnamefont {C.}~\bibnamefont {Martinez}}, \
  and\ \bibinfo {author} {\bibfnamefont {L.}~\bibnamefont {Pena}},\ }\href
  {\doibase 10.1088/0264-9381/11/11/014} {\bibfield  {journal} {\bibinfo
  {journal} {Class. Quant. Grav.}\ }\textbf {\bibinfo {volume} {11}},\ \bibinfo
  {pages} {2731} (\bibinfo {year} {1994})},\ \Eprint
  {http://arxiv.org/abs/gr-qc/9401025} {arXiv:gr-qc/9401025} \BibitemShut
  {NoStop}%
\bibitem [{\citenamefont {Cardoso}\ and\ \citenamefont
  {Lemos}(2001)}]{PhysRevD.64.084017}%
  \BibitemOpen
  \bibfield  {author} {\bibinfo {author} {\bibfnamefont {V.}~\bibnamefont
  {Cardoso}}\ and\ \bibinfo {author} {\bibfnamefont {J.~P.~S.}\ \bibnamefont
  {Lemos}},\ }\href {\doibase 10.1103/PhysRevD.64.084017} {\bibfield  {journal}
  {\bibinfo  {journal} {Phys. Rev. D}\ }\textbf {\bibinfo {volume} {64}},\
  \bibinfo {pages} {084017} (\bibinfo {year} {2001})}\BibitemShut {NoStop}%
\bibitem [{\citenamefont {Srednicki}(1999)}]{Srednicki_1999}%
  \BibitemOpen
  \bibfield  {author} {\bibinfo {author} {\bibfnamefont {M.}~\bibnamefont
  {Srednicki}},\ }\href {\doibase 10.1088/0305-4470/32/7/007} {\bibfield
  {journal} {\bibinfo  {journal} {Journal of Physics A: Mathematical and
  General}\ }\textbf {\bibinfo {volume} {32}},\ \bibinfo {pages} {1163–1175}
  (\bibinfo {year} {1999})}\BibitemShut {NoStop}%
\bibitem [{\citenamefont {D’Alessio}\ \emph {et~al.}(2016)\citenamefont
  {D’Alessio}, \citenamefont {Kafri}, \citenamefont {Polkovnikov},\ and\
  \citenamefont {Rigol}}]{D_Alessio_2016}%
  \BibitemOpen
  \bibfield  {author} {\bibinfo {author} {\bibfnamefont {L.}~\bibnamefont
  {D’Alessio}}, \bibinfo {author} {\bibfnamefont {Y.}~\bibnamefont {Kafri}},
  \bibinfo {author} {\bibfnamefont {A.}~\bibnamefont {Polkovnikov}}, \ and\
  \bibinfo {author} {\bibfnamefont {M.}~\bibnamefont {Rigol}},\ }\href
  {\doibase 10.1080/00018732.2016.1198134} {\bibfield  {journal} {\bibinfo
  {journal} {Advances in Physics}\ }\textbf {\bibinfo {volume} {65}},\ \bibinfo
  {pages} {239–362} (\bibinfo {year} {2016})}\BibitemShut {NoStop}%
\bibitem [{\citenamefont {Lashkari}\ \emph {et~al.}(2016)\citenamefont
  {Lashkari}, \citenamefont {Dymarsky},\ and\ \citenamefont
  {Liu}}]{lashkari2016eigenstatethermalizationhypothesisconformal}%
  \BibitemOpen
  \bibfield  {author} {\bibinfo {author} {\bibfnamefont {N.}~\bibnamefont
  {Lashkari}}, \bibinfo {author} {\bibfnamefont {A.}~\bibnamefont {Dymarsky}},
  \ and\ \bibinfo {author} {\bibfnamefont {H.}~\bibnamefont {Liu}},\ }\href
  {https://arxiv.org/abs/1610.00302} {\enquote {\bibinfo {title} {Eigenstate
  thermalization hypothesis in conformal field theory},}\ } (\bibinfo {year}
  {2016}),\ \Eprint {http://arxiv.org/abs/1610.00302} {arXiv:1610.00302
  [hep-th]} \BibitemShut {NoStop}%
\bibitem [{\citenamefont {Delacretaz}(2020)}]{10.21468/SciPostPhys.9.3.034}%
  \BibitemOpen
  \bibfield  {author} {\bibinfo {author} {\bibfnamefont {L.~V.}\ \bibnamefont
  {Delacretaz}},\ }\href {\doibase 10.21468/SciPostPhys.9.3.034} {\bibfield
  {journal} {\bibinfo  {journal} {SciPost Phys.}\ }\textbf {\bibinfo {volume}
  {9}},\ \bibinfo {pages} {034} (\bibinfo {year} {2020})}\BibitemShut {NoStop}%
\bibitem [{\citenamefont {Hellerman}\ \emph {et~al.}(2015)\citenamefont
  {Hellerman}, \citenamefont {Orlando}, \citenamefont {Reffert},\ and\
  \citenamefont {Watanabe}}]{Hellerman:2015nra}%
  \BibitemOpen
  \bibfield  {author} {\bibinfo {author} {\bibfnamefont {S.}~\bibnamefont
  {Hellerman}}, \bibinfo {author} {\bibfnamefont {D.}~\bibnamefont {Orlando}},
  \bibinfo {author} {\bibfnamefont {S.}~\bibnamefont {Reffert}}, \ and\
  \bibinfo {author} {\bibfnamefont {M.}~\bibnamefont {Watanabe}},\ }\href
  {\doibase 10.1007/JHEP12(2015)071} {\bibfield  {journal} {\bibinfo  {journal}
  {JHEP}\ }\textbf {\bibinfo {volume} {12}},\ \bibinfo {pages} {071} (\bibinfo
  {year} {2015})},\ \Eprint {http://arxiv.org/abs/1505.01537} {arXiv:1505.01537
  [hep-th]} \BibitemShut {NoStop}%
\bibitem [{\citenamefont {Alvarez-Gaume}\ \emph {et~al.}(2017)\citenamefont
  {Alvarez-Gaume}, \citenamefont {Loukas}, \citenamefont {Orlando},\ and\
  \citenamefont {Reffert}}]{Alvarez-Gaume:2016vff}%
  \BibitemOpen
  \bibfield  {author} {\bibinfo {author} {\bibfnamefont {L.}~\bibnamefont
  {Alvarez-Gaume}}, \bibinfo {author} {\bibfnamefont {O.}~\bibnamefont
  {Loukas}}, \bibinfo {author} {\bibfnamefont {D.}~\bibnamefont {Orlando}}, \
  and\ \bibinfo {author} {\bibfnamefont {S.}~\bibnamefont {Reffert}},\ }\href
  {\doibase 10.1007/JHEP04(2017)059} {\bibfield  {journal} {\bibinfo  {journal}
  {JHEP}\ }\textbf {\bibinfo {volume} {04}},\ \bibinfo {pages} {059} (\bibinfo
  {year} {2017})},\ \Eprint {http://arxiv.org/abs/1610.04495} {arXiv:1610.04495
  [hep-th]} \BibitemShut {NoStop}%
\bibitem [{\citenamefont {Orlando}\ \emph {et~al.}(2019)\citenamefont
  {Orlando}, \citenamefont {Reffert},\ and\ \citenamefont
  {Sannino}}]{Orlando:2019hte}%
  \BibitemOpen
  \bibfield  {author} {\bibinfo {author} {\bibfnamefont {D.}~\bibnamefont
  {Orlando}}, \bibinfo {author} {\bibfnamefont {S.}~\bibnamefont {Reffert}}, \
  and\ \bibinfo {author} {\bibfnamefont {F.}~\bibnamefont {Sannino}},\ }\href
  {\doibase 10.1007/JHEP08(2019)164} {\bibfield  {journal} {\bibinfo  {journal}
  {JHEP}\ }\textbf {\bibinfo {volume} {08}},\ \bibinfo {pages} {164} (\bibinfo
  {year} {2019})},\ \Eprint {http://arxiv.org/abs/1905.00026} {arXiv:1905.00026
  [hep-th]} \BibitemShut {NoStop}%
\bibitem [{\citenamefont {Gaum\'e}\ \emph {et~al.}(2021)\citenamefont
  {Gaum\'e}, \citenamefont {Orlando},\ and\ \citenamefont
  {Reffert}}]{Gaume:2020bmp}%
  \BibitemOpen
  \bibfield  {author} {\bibinfo {author} {\bibfnamefont {L.~A.}\ \bibnamefont
  {Gaum\'e}}, \bibinfo {author} {\bibfnamefont {D.}~\bibnamefont {Orlando}}, \
  and\ \bibinfo {author} {\bibfnamefont {S.}~\bibnamefont {Reffert}},\ }\href
  {\doibase 10.1016/j.physrep.2021.08.001} {\bibfield  {journal} {\bibinfo
  {journal} {Phys. Rept.}\ }\textbf {\bibinfo {volume} {933}},\ \bibinfo
  {pages} {1} (\bibinfo {year} {2021})},\ \Eprint
  {http://arxiv.org/abs/2008.03308} {arXiv:2008.03308 [hep-th]} \BibitemShut
  {NoStop}%
\bibitem [{\citenamefont {Nakayama}\ and\ \citenamefont
  {Nomura}(2015)}]{Nakayama:2015hga}%
  \BibitemOpen
  \bibfield  {author} {\bibinfo {author} {\bibfnamefont {Y.}~\bibnamefont
  {Nakayama}}\ and\ \bibinfo {author} {\bibfnamefont {Y.}~\bibnamefont
  {Nomura}},\ }\href {\doibase 10.1103/PhysRevD.92.126006} {\bibfield
  {journal} {\bibinfo  {journal} {Phys. Rev. D}\ }\textbf {\bibinfo {volume}
  {92}},\ \bibinfo {pages} {126006} (\bibinfo {year} {2015})},\ \Eprint
  {http://arxiv.org/abs/1509.01647} {arXiv:1509.01647 [hep-th]} \BibitemShut
  {NoStop}%
\bibitem [{\citenamefont {Aharony}\ and\ \citenamefont
  {Palti}(2021)}]{Aharony:2021mpc}%
  \BibitemOpen
  \bibfield  {author} {\bibinfo {author} {\bibfnamefont {O.}~\bibnamefont
  {Aharony}}\ and\ \bibinfo {author} {\bibfnamefont {E.}~\bibnamefont
  {Palti}},\ }\href {\doibase 10.1103/PhysRevD.104.126005} {\bibfield
  {journal} {\bibinfo  {journal} {Phys. Rev. D}\ }\textbf {\bibinfo {volume}
  {104}},\ \bibinfo {pages} {126005} (\bibinfo {year} {2021})},\ \Eprint
  {http://arxiv.org/abs/2108.04594} {arXiv:2108.04594 [hep-th]} \BibitemShut
  {NoStop}%
\bibitem [{\citenamefont {Benjamin}\ \emph {et~al.}(2016)\citenamefont
  {Benjamin}, \citenamefont {Dyer}, \citenamefont {Fitzpatrick},\ and\
  \citenamefont {Kachru}}]{Benjamin:2016fhe}%
  \BibitemOpen
  \bibfield  {author} {\bibinfo {author} {\bibfnamefont {N.}~\bibnamefont
  {Benjamin}}, \bibinfo {author} {\bibfnamefont {E.}~\bibnamefont {Dyer}},
  \bibinfo {author} {\bibfnamefont {A.~L.}\ \bibnamefont {Fitzpatrick}}, \ and\
  \bibinfo {author} {\bibfnamefont {S.}~\bibnamefont {Kachru}},\ }\href
  {\doibase 10.1007/JHEP08(2016)041} {\bibfield  {journal} {\bibinfo  {journal}
  {JHEP}\ }\textbf {\bibinfo {volume} {08}},\ \bibinfo {pages} {041} (\bibinfo
  {year} {2016})},\ \Eprint {http://arxiv.org/abs/1603.09745} {arXiv:1603.09745
  [hep-th]} \BibitemShut {NoStop}%
\bibitem [{\citenamefont {Crisford}\ \emph {et~al.}(2018)\citenamefont
  {Crisford}, \citenamefont {Horowitz},\ and\ \citenamefont
  {Santos}}]{Crisford:2017gsb}%
  \BibitemOpen
  \bibfield  {author} {\bibinfo {author} {\bibfnamefont {T.}~\bibnamefont
  {Crisford}}, \bibinfo {author} {\bibfnamefont {G.~T.}\ \bibnamefont
  {Horowitz}}, \ and\ \bibinfo {author} {\bibfnamefont {J.~E.}\ \bibnamefont
  {Santos}},\ }\href {\doibase 10.1103/PhysRevD.97.066005} {\bibfield
  {journal} {\bibinfo  {journal} {Phys. Rev. D}\ }\textbf {\bibinfo {volume}
  {97}},\ \bibinfo {pages} {066005} (\bibinfo {year} {2018})},\ \Eprint
  {http://arxiv.org/abs/1709.07880} {arXiv:1709.07880 [hep-th]} \BibitemShut
  {NoStop}%
\bibitem [{\citenamefont {Montero}(2019)}]{Montero:2018fns}%
  \BibitemOpen
  \bibfield  {author} {\bibinfo {author} {\bibfnamefont {M.}~\bibnamefont
  {Montero}},\ }\href {\doibase 10.1007/JHEP03(2019)157} {\bibfield  {journal}
  {\bibinfo  {journal} {JHEP}\ }\textbf {\bibinfo {volume} {03}},\ \bibinfo
  {pages} {157} (\bibinfo {year} {2019})},\ \Eprint
  {http://arxiv.org/abs/1812.03978} {arXiv:1812.03978 [hep-th]} \BibitemShut
  {NoStop}%
\bibitem [{\citenamefont {Cremonini}\ \emph {et~al.}(2010)\citenamefont
  {Cremonini}, \citenamefont {Liu},\ and\ \citenamefont
  {Szepietowski}}]{Cremonini:2009ih}%
  \BibitemOpen
  \bibfield  {author} {\bibinfo {author} {\bibfnamefont {S.}~\bibnamefont
  {Cremonini}}, \bibinfo {author} {\bibfnamefont {J.~T.}\ \bibnamefont {Liu}},
  \ and\ \bibinfo {author} {\bibfnamefont {P.}~\bibnamefont {Szepietowski}},\
  }\href {\doibase 10.1007/JHEP03(2010)042} {\bibfield  {journal} {\bibinfo
  {journal} {JHEP}\ }\textbf {\bibinfo {volume} {03}},\ \bibinfo {pages} {042}
  (\bibinfo {year} {2010})},\ \Eprint {http://arxiv.org/abs/0910.5159}
  {arXiv:0910.5159 [hep-th]} \BibitemShut {NoStop}%
\bibitem [{\citenamefont {Boulware}\ and\ \citenamefont
  {Deser}(1985)}]{PhysRevLett.55.2656}%
  \BibitemOpen
  \bibfield  {author} {\bibinfo {author} {\bibfnamefont {D.~G.}\ \bibnamefont
  {Boulware}}\ and\ \bibinfo {author} {\bibfnamefont {S.}~\bibnamefont
  {Deser}},\ }\href {\doibase 10.1103/PhysRevLett.55.2656} {\bibfield
  {journal} {\bibinfo  {journal} {Phys. Rev. Lett.}\ }\textbf {\bibinfo
  {volume} {55}},\ \bibinfo {pages} {2656} (\bibinfo {year}
  {1985})}\BibitemShut {NoStop}%
\bibitem [{\citenamefont {Nojiri}\ \emph {et~al.}(2001)\citenamefont {Nojiri},
  \citenamefont {Odintsov},\ and\ \citenamefont {Ogushi}}]{Nojiri_2001}%
  \BibitemOpen
  \bibfield  {author} {\bibinfo {author} {\bibfnamefont {S.}~\bibnamefont
  {Nojiri}}, \bibinfo {author} {\bibfnamefont {S.~D.}\ \bibnamefont
  {Odintsov}}, \ and\ \bibinfo {author} {\bibfnamefont {S.}~\bibnamefont
  {Ogushi}},\ }\href {\doibase 10.1103/physrevd.65.023521} {\bibfield
  {journal} {\bibinfo  {journal} {Physical Review D}\ }\textbf {\bibinfo
  {volume} {65}} (\bibinfo {year} {2001}),\
  10.1103/physrevd.65.023521}\BibitemShut {NoStop}%
\bibitem [{\citenamefont {Nojiri}\ and\ \citenamefont
  {Odintsov}(2006)}]{NOJIRI2006144}%
  \BibitemOpen
  \bibfield  {author} {\bibinfo {author} {\bibfnamefont {S.}~\bibnamefont
  {Nojiri}}\ and\ \bibinfo {author} {\bibfnamefont {S.~D.}\ \bibnamefont
  {Odintsov}},\ }\href {\doibase
  https://doi.org/10.1016/j.physletb.2006.06.065} {\bibfield  {journal}
  {\bibinfo  {journal} {Physics Letters B}\ }\textbf {\bibinfo {volume}
  {639}},\ \bibinfo {pages} {144} (\bibinfo {year} {2006})}\BibitemShut
  {NoStop}%
\bibitem [{\citenamefont {Cai}(2002)}]{Cai:2001dz}%
  \BibitemOpen
  \bibfield  {author} {\bibinfo {author} {\bibfnamefont {R.-G.}\ \bibnamefont
  {Cai}},\ }\href {\doibase 10.1103/PhysRevD.65.084014} {\bibfield  {journal}
  {\bibinfo  {journal} {Phys. Rev. D}\ }\textbf {\bibinfo {volume} {65}},\
  \bibinfo {pages} {084014} (\bibinfo {year} {2002})},\ \Eprint
  {http://arxiv.org/abs/hep-th/0109133} {arXiv:hep-th/0109133} \BibitemShut
  {NoStop}%
\bibitem [{\citenamefont {Cvetič}\ \emph {et~al.}(2002)\citenamefont
  {Cvetič}, \citenamefont {Nojiri},\ and\ \citenamefont
  {Odintsov}}]{Cveti__2002}%
  \BibitemOpen
  \bibfield  {author} {\bibinfo {author} {\bibfnamefont {M.}~\bibnamefont
  {Cvetič}}, \bibinfo {author} {\bibfnamefont {S.}~\bibnamefont {Nojiri}}, \
  and\ \bibinfo {author} {\bibfnamefont {S.}~\bibnamefont {Odintsov}},\ }\href
  {\doibase 10.1016/s0550-3213(02)00075-5} {\bibfield  {journal} {\bibinfo
  {journal} {Nuclear Physics B}\ }\textbf {\bibinfo {volume} {628}},\ \bibinfo
  {pages} {295–330} (\bibinfo {year} {2002})}\BibitemShut {NoStop}%
\bibitem [{\citenamefont {Cai}\ \emph {et~al.}(2013)\citenamefont {Cai},
  \citenamefont {Cao}, \citenamefont {Li},\ and\ \citenamefont
  {Yang}}]{Cai:2013qga}%
  \BibitemOpen
  \bibfield  {author} {\bibinfo {author} {\bibfnamefont {R.-G.}\ \bibnamefont
  {Cai}}, \bibinfo {author} {\bibfnamefont {L.-M.}\ \bibnamefont {Cao}},
  \bibinfo {author} {\bibfnamefont {L.}~\bibnamefont {Li}}, \ and\ \bibinfo
  {author} {\bibfnamefont {R.-Q.}\ \bibnamefont {Yang}},\ }\href {\doibase
  10.1007/JHEP09(2013)005} {\bibfield  {journal} {\bibinfo  {journal} {JHEP}\
  }\textbf {\bibinfo {volume} {09}},\ \bibinfo {pages} {005} (\bibinfo {year}
  {2013})},\ \Eprint {http://arxiv.org/abs/1306.6233} {arXiv:1306.6233 [gr-qc]}
  \BibitemShut {NoStop}%
\bibitem [{\citenamefont {Cai}\ \emph {et~al.}(2010)\citenamefont {Cai},
  \citenamefont {Liu},\ and\ \citenamefont {Sun}}]{Cai:2009zn}%
  \BibitemOpen
  \bibfield  {author} {\bibinfo {author} {\bibfnamefont {R.-G.}\ \bibnamefont
  {Cai}}, \bibinfo {author} {\bibfnamefont {Y.}~\bibnamefont {Liu}}, \ and\
  \bibinfo {author} {\bibfnamefont {Y.-W.}\ \bibnamefont {Sun}},\ }\href
  {\doibase 10.1007/JHEP04(2010)090} {\bibfield  {journal} {\bibinfo  {journal}
  {JHEP}\ }\textbf {\bibinfo {volume} {04}},\ \bibinfo {pages} {090} (\bibinfo
  {year} {2010})},\ \Eprint {http://arxiv.org/abs/0910.4705} {arXiv:0910.4705
  [hep-th]} \BibitemShut {NoStop}%
\bibitem [{\citenamefont {Cai}\ \emph {et~al.}(2011)\citenamefont {Cai},
  \citenamefont {Nie},\ and\ \citenamefont {Zhang}}]{Cai:2010zm}%
  \BibitemOpen
  \bibfield  {author} {\bibinfo {author} {\bibfnamefont {R.-G.}\ \bibnamefont
  {Cai}}, \bibinfo {author} {\bibfnamefont {Z.-Y.}\ \bibnamefont {Nie}}, \ and\
  \bibinfo {author} {\bibfnamefont {H.-Q.}\ \bibnamefont {Zhang}},\ }\href
  {\doibase 10.1103/PhysRevD.83.066013} {\bibfield  {journal} {\bibinfo
  {journal} {Phys. Rev. D}\ }\textbf {\bibinfo {volume} {83}},\ \bibinfo
  {pages} {066013} (\bibinfo {year} {2011})},\ \Eprint
  {http://arxiv.org/abs/1012.5559} {arXiv:1012.5559 [hep-th]} \BibitemShut
  {NoStop}%
\bibitem [{\citenamefont {Camanho}\ \emph {et~al.}(2016)\citenamefont
  {Camanho}, \citenamefont {Edelstein}, \citenamefont {Maldacena},\ and\
  \citenamefont {Zhiboedov}}]{Camanho_2016}%
  \BibitemOpen
  \bibfield  {author} {\bibinfo {author} {\bibfnamefont {X.~O.}\ \bibnamefont
  {Camanho}}, \bibinfo {author} {\bibfnamefont {J.~D.}\ \bibnamefont
  {Edelstein}}, \bibinfo {author} {\bibfnamefont {J.}~\bibnamefont
  {Maldacena}}, \ and\ \bibinfo {author} {\bibfnamefont {A.}~\bibnamefont
  {Zhiboedov}},\ }\href {\doibase 10.1007/jhep02(2016)020} {\bibfield
  {journal} {\bibinfo  {journal} {Journal of High Energy Physics}\ }\textbf
  {\bibinfo {volume} {2016}} (\bibinfo {year} {2016}),\
  10.1007/jhep02(2016)020}\BibitemShut {NoStop}%
\bibitem [{\citenamefont {Afkhami-Jeddi}\ \emph
  {et~al.}(2017{\natexlab{a}})\citenamefont {Afkhami-Jeddi}, \citenamefont
  {Hartman}, \citenamefont {Kundu},\ and\ \citenamefont
  {Tajdini}}]{Afkhami_Jeddi_2017}%
  \BibitemOpen
  \bibfield  {author} {\bibinfo {author} {\bibfnamefont {N.}~\bibnamefont
  {Afkhami-Jeddi}}, \bibinfo {author} {\bibfnamefont {T.}~\bibnamefont
  {Hartman}}, \bibinfo {author} {\bibfnamefont {S.}~\bibnamefont {Kundu}}, \
  and\ \bibinfo {author} {\bibfnamefont {A.}~\bibnamefont {Tajdini}},\ }\href
  {\doibase 10.1007/jhep12(2017)049} {\bibfield  {journal} {\bibinfo  {journal}
  {Journal of High Energy Physics}\ }\textbf {\bibinfo {volume} {2017}}
  (\bibinfo {year} {2017}{\natexlab{a}}),\ 10.1007/jhep12(2017)049}\BibitemShut
  {NoStop}%
\bibitem [{\citenamefont {Afkhami-Jeddi}\ \emph
  {et~al.}(2017{\natexlab{b}})\citenamefont {Afkhami-Jeddi}, \citenamefont
  {Hartman}, \citenamefont {Kundu},\ and\ \citenamefont
  {Tajdini}}]{afkhamijeddi2017shockwavesoperatorproductexpansion}%
  \BibitemOpen
  \bibfield  {author} {\bibinfo {author} {\bibfnamefont {N.}~\bibnamefont
  {Afkhami-Jeddi}}, \bibinfo {author} {\bibfnamefont {T.}~\bibnamefont
  {Hartman}}, \bibinfo {author} {\bibfnamefont {S.}~\bibnamefont {Kundu}}, \
  and\ \bibinfo {author} {\bibfnamefont {A.}~\bibnamefont {Tajdini}},\ }\href
  {\doibase 10.1007/JHEP03(2019)201} {\enquote {\bibinfo {title} {Shockwaves
  from the operator product expansion},}\ } (\bibinfo {year}
  {2017}{\natexlab{b}})\BibitemShut {NoStop}%
\bibitem [{\citenamefont {Kulaxizi}\ \emph {et~al.}(2018)\citenamefont
  {Kulaxizi}, \citenamefont {Parnachev},\ and\ \citenamefont
  {Zhiboedov}}]{Kulaxizi_2018}%
  \BibitemOpen
  \bibfield  {author} {\bibinfo {author} {\bibfnamefont {M.}~\bibnamefont
  {Kulaxizi}}, \bibinfo {author} {\bibfnamefont {A.}~\bibnamefont {Parnachev}},
  \ and\ \bibinfo {author} {\bibfnamefont {A.}~\bibnamefont {Zhiboedov}},\
  }\href {\doibase 10.1007/jhep06(2018)121} {\bibfield  {journal} {\bibinfo
  {journal} {Journal of High Energy Physics}\ }\textbf {\bibinfo {volume}
  {2018}} (\bibinfo {year} {2018}),\ 10.1007/jhep06(2018)121}\BibitemShut
  {NoStop}%
\bibitem [{\citenamefont {Costa}\ \emph {et~al.}(2017)\citenamefont {Costa},
  \citenamefont {Hansen},\ and\ \citenamefont {Penedones}}]{Costa_2017}%
  \BibitemOpen
  \bibfield  {author} {\bibinfo {author} {\bibfnamefont {M.~S.}\ \bibnamefont
  {Costa}}, \bibinfo {author} {\bibfnamefont {T.}~\bibnamefont {Hansen}}, \
  and\ \bibinfo {author} {\bibfnamefont {J.}~\bibnamefont {Penedones}},\ }\href
  {\doibase 10.1007/jhep10(2017)197} {\bibfield  {journal} {\bibinfo  {journal}
  {Journal of High Energy Physics}\ }\textbf {\bibinfo {volume} {2017}}
  (\bibinfo {year} {2017}),\ 10.1007/jhep10(2017)197}\BibitemShut {NoStop}%
\bibitem [{\citenamefont {Caron-Huot}\ \emph {et~al.}(2023)\citenamefont
  {Caron-Huot}, \citenamefont {Li}, \citenamefont {Parra-Martinez},\ and\
  \citenamefont {Simmons-Duffin}}]{Caron_Huot_2023}%
  \BibitemOpen
  \bibfield  {author} {\bibinfo {author} {\bibfnamefont {S.}~\bibnamefont
  {Caron-Huot}}, \bibinfo {author} {\bibfnamefont {Y.-Z.}\ \bibnamefont {Li}},
  \bibinfo {author} {\bibfnamefont {J.}~\bibnamefont {Parra-Martinez}}, \ and\
  \bibinfo {author} {\bibfnamefont {D.}~\bibnamefont {Simmons-Duffin}},\ }\href
  {\doibase 10.1007/jhep05(2023)122} {\bibfield  {journal} {\bibinfo  {journal}
  {Journal of High Energy Physics}\ }\textbf {\bibinfo {volume} {2023}}
  (\bibinfo {year} {2023}),\ 10.1007/jhep05(2023)122}\BibitemShut {NoStop}%
\bibitem [{\citenamefont {Grinberg}\ and\ \citenamefont
  {Maldacena}(2021)}]{Grinberg_2021}%
  \BibitemOpen
  \bibfield  {author} {\bibinfo {author} {\bibfnamefont {M.}~\bibnamefont
  {Grinberg}}\ and\ \bibinfo {author} {\bibfnamefont {J.}~\bibnamefont
  {Maldacena}},\ }\href {\doibase 10.1007/jhep03(2021)131} {\bibfield
  {journal} {\bibinfo  {journal} {Journal of High Energy Physics}\ }\textbf
  {\bibinfo {volume} {2021}} (\bibinfo {year} {2021}),\
  10.1007/jhep03(2021)131}\BibitemShut {NoStop}%
\bibitem [{\citenamefont {Alday}\ and\ \citenamefont
  {Maldacena}(2007)}]{Alday:2007mf}%
  \BibitemOpen
  \bibfield  {author} {\bibinfo {author} {\bibfnamefont {L.~F.}\ \bibnamefont
  {Alday}}\ and\ \bibinfo {author} {\bibfnamefont {J.~M.}\ \bibnamefont
  {Maldacena}},\ }\href {\doibase 10.1088/1126-6708/2007/11/019} {\bibfield
  {journal} {\bibinfo  {journal} {JHEP}\ }\textbf {\bibinfo {volume} {11}},\
  \bibinfo {pages} {019} (\bibinfo {year} {2007})},\ \Eprint
  {http://arxiv.org/abs/0708.0672} {arXiv:0708.0672 [hep-th]} \BibitemShut
  {NoStop}%
\bibitem [{\citenamefont {Komargodski}\ and\ \citenamefont
  {Zhiboedov}(2013)}]{Komargodski:2012ek}%
  \BibitemOpen
  \bibfield  {author} {\bibinfo {author} {\bibfnamefont {Z.}~\bibnamefont
  {Komargodski}}\ and\ \bibinfo {author} {\bibfnamefont {A.}~\bibnamefont
  {Zhiboedov}},\ }\href {\doibase 10.1007/JHEP11(2013)140} {\bibfield
  {journal} {\bibinfo  {journal} {JHEP}\ }\textbf {\bibinfo {volume} {11}},\
  \bibinfo {pages} {140} (\bibinfo {year} {2013})},\ \Eprint
  {http://arxiv.org/abs/1212.4103} {arXiv:1212.4103 [hep-th]} \BibitemShut
  {NoStop}%
\bibitem [{\citenamefont {Alday}\ \emph {et~al.}(2015)\citenamefont {Alday},
  \citenamefont {Bissi},\ and\ \citenamefont {Lukowski}}]{Alday:2015eya}%
  \BibitemOpen
  \bibfield  {author} {\bibinfo {author} {\bibfnamefont {L.~F.}\ \bibnamefont
  {Alday}}, \bibinfo {author} {\bibfnamefont {A.}~\bibnamefont {Bissi}}, \ and\
  \bibinfo {author} {\bibfnamefont {T.}~\bibnamefont {Lukowski}},\ }\href
  {\doibase 10.1007/JHEP11(2015)101} {\bibfield  {journal} {\bibinfo  {journal}
  {JHEP}\ }\textbf {\bibinfo {volume} {11}},\ \bibinfo {pages} {101} (\bibinfo
  {year} {2015})},\ \Eprint {http://arxiv.org/abs/1502.07707} {arXiv:1502.07707
  [hep-th]} \BibitemShut {NoStop}%
\bibitem [{\citenamefont {Kulaxizi}\ \emph {et~al.}(2019)\citenamefont
  {Kulaxizi}, \citenamefont {Ng},\ and\ \citenamefont
  {Parnachev}}]{Kulaxizi:2018dxo}%
  \BibitemOpen
  \bibfield  {author} {\bibinfo {author} {\bibfnamefont {M.}~\bibnamefont
  {Kulaxizi}}, \bibinfo {author} {\bibfnamefont {G.~S.}\ \bibnamefont {Ng}}, \
  and\ \bibinfo {author} {\bibfnamefont {A.}~\bibnamefont {Parnachev}},\ }\href
  {\doibase 10.21468/SciPostPhys.6.6.065} {\bibfield  {journal} {\bibinfo
  {journal} {SciPost Phys.}\ }\textbf {\bibinfo {volume} {6}},\ \bibinfo
  {pages} {065} (\bibinfo {year} {2019})},\ \Eprint
  {http://arxiv.org/abs/1812.03120} {arXiv:1812.03120 [hep-th]} \BibitemShut
  {NoStop}%
\bibitem [{\citenamefont {Karlsson}\ \emph {et~al.}(2019)\citenamefont
  {Karlsson}, \citenamefont {Kulaxizi}, \citenamefont {Parnachev},\ and\
  \citenamefont {Tadi\'c}}]{Karlsson:2019qfi}%
  \BibitemOpen
  \bibfield  {author} {\bibinfo {author} {\bibfnamefont {R.}~\bibnamefont
  {Karlsson}}, \bibinfo {author} {\bibfnamefont {M.}~\bibnamefont {Kulaxizi}},
  \bibinfo {author} {\bibfnamefont {A.}~\bibnamefont {Parnachev}}, \ and\
  \bibinfo {author} {\bibfnamefont {P.}~\bibnamefont {Tadi\'c}},\ }\href
  {\doibase 10.1007/JHEP10(2019)046} {\bibfield  {journal} {\bibinfo  {journal}
  {JHEP}\ }\textbf {\bibinfo {volume} {10}},\ \bibinfo {pages} {046} (\bibinfo
  {year} {2019})},\ \Eprint {http://arxiv.org/abs/1904.00060} {arXiv:1904.00060
  [hep-th]} \BibitemShut {NoStop}%
\bibitem [{\citenamefont {Li}(2020)}]{Li:2019zba}%
  \BibitemOpen
  \bibfield  {author} {\bibinfo {author} {\bibfnamefont {Y.-Z.}\ \bibnamefont
  {Li}},\ }\href {\doibase 10.1007/JHEP07(2020)046} {\bibfield  {journal}
  {\bibinfo  {journal} {JHEP}\ }\textbf {\bibinfo {volume} {07}},\ \bibinfo
  {pages} {046} (\bibinfo {year} {2020})},\ \Eprint
  {http://arxiv.org/abs/1910.06357} {arXiv:1910.06357 [hep-th]} \BibitemShut
  {NoStop}%
\bibitem [{\citenamefont {Li}\ and\ \citenamefont {Zhang}(2020)}]{Li:2020dqm}%
  \BibitemOpen
  \bibfield  {author} {\bibinfo {author} {\bibfnamefont {Y.-Z.}\ \bibnamefont
  {Li}}\ and\ \bibinfo {author} {\bibfnamefont {H.-Y.}\ \bibnamefont {Zhang}},\
  }\href {\doibase 10.1007/JHEP10(2020)055} {\bibfield  {journal} {\bibinfo
  {journal} {JHEP}\ }\textbf {\bibinfo {volume} {10}},\ \bibinfo {pages} {055}
  (\bibinfo {year} {2020})},\ \Eprint {http://arxiv.org/abs/2004.04758}
  {arXiv:2004.04758 [hep-th]} \BibitemShut {NoStop}%
\bibitem [{\citenamefont {Brigante}\ \emph {et~al.}(2008)\citenamefont
  {Brigante}, \citenamefont {Liu}, \citenamefont {Myers}, \citenamefont
  {Shenker},\ and\ \citenamefont {Yaida}}]{PhysRevD.77.126006}%
  \BibitemOpen
  \bibfield  {author} {\bibinfo {author} {\bibfnamefont {M.}~\bibnamefont
  {Brigante}}, \bibinfo {author} {\bibfnamefont {H.}~\bibnamefont {Liu}},
  \bibinfo {author} {\bibfnamefont {R.~C.}\ \bibnamefont {Myers}}, \bibinfo
  {author} {\bibfnamefont {S.}~\bibnamefont {Shenker}}, \ and\ \bibinfo
  {author} {\bibfnamefont {S.}~\bibnamefont {Yaida}},\ }\href {\doibase
  10.1103/PhysRevD.77.126006} {\bibfield  {journal} {\bibinfo  {journal} {Phys.
  Rev. D}\ }\textbf {\bibinfo {volume} {77}},\ \bibinfo {pages} {126006}
  (\bibinfo {year} {2008})}\BibitemShut {NoStop}%
\bibitem [{\citenamefont {Pugliese}\ \emph
  {et~al.}(2011{\natexlab{a}})\citenamefont {Pugliese}, \citenamefont
  {Quevedo},\ and\ \citenamefont {Ruffini}}]{Pugliese:2010ps}%
  \BibitemOpen
  \bibfield  {author} {\bibinfo {author} {\bibfnamefont {D.}~\bibnamefont
  {Pugliese}}, \bibinfo {author} {\bibfnamefont {H.}~\bibnamefont {Quevedo}}, \
  and\ \bibinfo {author} {\bibfnamefont {R.}~\bibnamefont {Ruffini}},\ }\href
  {\doibase 10.1103/PhysRevD.83.024021} {\bibfield  {journal} {\bibinfo
  {journal} {Phys. Rev. D}\ }\textbf {\bibinfo {volume} {83}},\ \bibinfo
  {pages} {024021} (\bibinfo {year} {2011}{\natexlab{a}})},\ \Eprint
  {http://arxiv.org/abs/1012.5411} {arXiv:1012.5411 [astro-ph.HE]} \BibitemShut
  {NoStop}%
\bibitem [{\citenamefont {Pugliese}\ \emph
  {et~al.}(2011{\natexlab{b}})\citenamefont {Pugliese}, \citenamefont
  {Quevedo},\ and\ \citenamefont {Ruffini}}]{Pugliese:2011py}%
  \BibitemOpen
  \bibfield  {author} {\bibinfo {author} {\bibfnamefont {D.}~\bibnamefont
  {Pugliese}}, \bibinfo {author} {\bibfnamefont {H.}~\bibnamefont {Quevedo}}, \
  and\ \bibinfo {author} {\bibfnamefont {R.}~\bibnamefont {Ruffini}},\ }\href
  {\doibase 10.1103/PhysRevD.83.104052} {\bibfield  {journal} {\bibinfo
  {journal} {Phys. Rev. D}\ }\textbf {\bibinfo {volume} {83}},\ \bibinfo
  {pages} {104052} (\bibinfo {year} {2011}{\natexlab{b}})},\ \Eprint
  {http://arxiv.org/abs/1103.1807} {arXiv:1103.1807 [gr-qc]} \BibitemShut
  {NoStop}%
\bibitem [{\citenamefont {Pugliese}\ \emph {et~al.}(2017)\citenamefont
  {Pugliese}, \citenamefont {Quevedo},\ and\ \citenamefont
  {Ruffini}}]{Pugliese:2013xfa}%
  \BibitemOpen
  \bibfield  {author} {\bibinfo {author} {\bibfnamefont {D.}~\bibnamefont
  {Pugliese}}, \bibinfo {author} {\bibfnamefont {H.}~\bibnamefont {Quevedo}}, \
  and\ \bibinfo {author} {\bibfnamefont {R.}~\bibnamefont {Ruffini}},\ }\href
  {\doibase 10.1140/epjc/s10052-017-4769-x} {\bibfield  {journal} {\bibinfo
  {journal} {Eur. Phys. J. C}\ }\textbf {\bibinfo {volume} {77}},\ \bibinfo
  {pages} {206} (\bibinfo {year} {2017})},\ \Eprint
  {http://arxiv.org/abs/1304.2940} {arXiv:1304.2940 [gr-qc]} \BibitemShut
  {NoStop}%
\bibitem [{\citenamefont {Chandrasekhar}\ and\ \citenamefont
  {Mohapatra}(2019)}]{Chandrasekhar:2018sjg}%
  \BibitemOpen
  \bibfield  {author} {\bibinfo {author} {\bibfnamefont {B.}~\bibnamefont
  {Chandrasekhar}}\ and\ \bibinfo {author} {\bibfnamefont {S.}~\bibnamefont
  {Mohapatra}},\ }\href {\doibase 10.1016/j.physletb.2019.02.042} {\bibfield
  {journal} {\bibinfo  {journal} {Phys. Lett. B}\ }\textbf {\bibinfo {volume}
  {791}},\ \bibinfo {pages} {367} (\bibinfo {year} {2019})},\ \Eprint
  {http://arxiv.org/abs/1805.05088} {arXiv:1805.05088 [hep-th]} \BibitemShut
  {NoStop}%
\bibitem [{\citenamefont {Guo}\ and\ \citenamefont {Li}(2020)}]{Guo:2020zmf}%
  \BibitemOpen
  \bibfield  {author} {\bibinfo {author} {\bibfnamefont {M.}~\bibnamefont
  {Guo}}\ and\ \bibinfo {author} {\bibfnamefont {P.-C.}\ \bibnamefont {Li}},\
  }\href {\doibase 10.1140/epjc/s10052-020-8164-7} {\bibfield  {journal}
  {\bibinfo  {journal} {Eur. Phys. J. C}\ }\textbf {\bibinfo {volume} {80}},\
  \bibinfo {pages} {588} (\bibinfo {year} {2020})},\ \Eprint
  {http://arxiv.org/abs/2003.02523} {arXiv:2003.02523 [gr-qc]} \BibitemShut
  {NoStop}%
\bibitem [{\citenamefont {Jafferis}\ \emph {et~al.}(2018)\citenamefont
  {Jafferis}, \citenamefont {Mukhametzhanov},\ and\ \citenamefont
  {Zhiboedov}}]{Jafferis:2017zna}%
  \BibitemOpen
  \bibfield  {author} {\bibinfo {author} {\bibfnamefont {D.}~\bibnamefont
  {Jafferis}}, \bibinfo {author} {\bibfnamefont {B.}~\bibnamefont
  {Mukhametzhanov}}, \ and\ \bibinfo {author} {\bibfnamefont {A.}~\bibnamefont
  {Zhiboedov}},\ }\href {\doibase 10.1007/JHEP05(2018)043} {\bibfield
  {journal} {\bibinfo  {journal} {JHEP}\ }\textbf {\bibinfo {volume} {05}},\
  \bibinfo {pages} {043} (\bibinfo {year} {2018})},\ \Eprint
  {http://arxiv.org/abs/1710.11161} {arXiv:1710.11161 [hep-th]} \BibitemShut
  {NoStop}%
\bibitem [{\citenamefont {Breitenlohner}\ and\ \citenamefont
  {Freedman}(1982)}]{Breitenlohner:1982jf}%
  \BibitemOpen
  \bibfield  {author} {\bibinfo {author} {\bibfnamefont {P.}~\bibnamefont
  {Breitenlohner}}\ and\ \bibinfo {author} {\bibfnamefont {D.~Z.}\ \bibnamefont
  {Freedman}},\ }\href {\doibase 10.1016/0003-4916(82)90116-6} {\bibfield
  {journal} {\bibinfo  {journal} {Annals Phys.}\ }\textbf {\bibinfo {volume}
  {144}},\ \bibinfo {pages} {249} (\bibinfo {year} {1982})}\BibitemShut
  {NoStop}%
\bibitem [{\citenamefont {Hartnoll}\ \emph {et~al.}(2008)\citenamefont
  {Hartnoll}, \citenamefont {Herzog},\ and\ \citenamefont
  {Horowitz}}]{Hartnoll:2008vx}%
  \BibitemOpen
  \bibfield  {author} {\bibinfo {author} {\bibfnamefont {S.~A.}\ \bibnamefont
  {Hartnoll}}, \bibinfo {author} {\bibfnamefont {C.~P.}\ \bibnamefont
  {Herzog}}, \ and\ \bibinfo {author} {\bibfnamefont {G.~T.}\ \bibnamefont
  {Horowitz}},\ }\href {\doibase 10.1103/PhysRevLett.101.031601} {\bibfield
  {journal} {\bibinfo  {journal} {Phys. Rev. Lett.}\ }\textbf {\bibinfo
  {volume} {101}},\ \bibinfo {pages} {031601} (\bibinfo {year} {2008})},\
  \Eprint {http://arxiv.org/abs/0803.3295} {arXiv:0803.3295 [hep-th]}
  \BibitemShut {NoStop}%
\bibitem [{\citenamefont {Gubser}(2008)}]{Gubser:2008px}%
  \BibitemOpen
  \bibfield  {author} {\bibinfo {author} {\bibfnamefont {S.~S.}\ \bibnamefont
  {Gubser}},\ }\href {\doibase 10.1103/PhysRevD.78.065034} {\bibfield
  {journal} {\bibinfo  {journal} {Phys. Rev. D}\ }\textbf {\bibinfo {volume}
  {78}},\ \bibinfo {pages} {065034} (\bibinfo {year} {2008})},\ \Eprint
  {http://arxiv.org/abs/0801.2977} {arXiv:0801.2977 [hep-th]} \BibitemShut
  {NoStop}%
\end{thebibliography}%
\end{document}